\documentstyle[preprint,aps,tighten]{revtex}
\begin{document}

\draft
\title{Nonextensive statistics: Theoretical, experimental and computational evidences and connections.}
\author{Constantino Tsallis}
\address{Centro Brasileiro de Pesquisas F\'{\i}sicas\\
Rua Xavier Sigaud 150, 22290-180 Rio de Janeiro-RJ, Brazil\\
e-mail: tsallis@cbpf.br}
\maketitle

\begin{abstract}
The domain of validity of standard thermodynamics and Boltzmann-Gibbs statistical mechanics is 
discussed and then formally enlarged  in order to hopefully cover a variety of anomalous systems.  The 
generalization concerns {\it nonextensive} systems, where nonextensivity is understood in the 
thermodynamical sense. This generalization was first proposed in 1988 inspired by the probabilistic 
description of multifractal geometries, and has been intensively studied during this decade. In the present 
effort, after introducing some historical background, we briefly describe the formalism, and then exhibit 
the present status in what concerns theoretical, experimental and computational evidences and 
connections, as well as  some perspectives for the future. In addition to these, here and there we point out 
various (possibly!) relevant questions, whose answer would certainly clarify our current understanding of 
the foundations of statistical mechanics and its thermodynamical implications.\\
\end{abstract}

\narrowtext

\section{INTRODUCTION}

A diffuse belief exists, among many physicists as well as other scientists, that Boltzmann-Gibbs (BG) 
statistical mechanics and standard thermodynamics are {\it eternal} and {\it universal}. It is certainly fair 
to say that ``eternal", in precisely the same sense that Newtonian mechanics is ``eternal", they indeed are. 
But, again in complete analogy with Newtonian mechanics, we can by no means consider them as 
universal. Indeed, we all know that, when the involved velocities approach that of light, Newtonian 
mechanics becomes only an approximation (an increasingly bad one!) and reality is better described by 
special relativity. Analogously, when the involved masses are as small as say the electron mass, once 
again Newtonian mechanics becomes but a (bad) approximation, and quantum mechanics becomes 
necessary to understand nature. Also, if the involved masses are very large, Newtonian mechanics has to 
be extended  into general relativity.
In these senses we certainly cannot consider Newtonian mechanics as being universal. I believe that the 
same type of considerations apply to standard statistical mechanics and thermodynamics. Indeed, after 
more than one century highly successful applications of the magnificent Boltzmann's connection of 
Clausius {\it macroscopic} entropy to the theory of probabilities applied to the {\it microscopic} world , 
BG thermal statistics can (and should!) easily be considered as one of the pillars of modern science. 
However, it is {\it unavoidable} to think that, like all other products of human mind, this formalism {\it 
must} have physical restrictions, i.e., domains of applicability, out of which it can at best be but an 
approximation. It seems that BG statistics satisfactorily describe nature {\it if} the effective microscopic 
interactions are {\it short}-ranged (i.e., {\it close} spatial connections) {\it and} the effective microscopic 
memory is {\it short}-ranged (i.e., {\it close} time connections) {\it and} the boundary conditions are {\it 
non(multi)fractal}. Roughly
speaking, the standard formalisms are applicable whenever (and probably only whenever) the relevant 
space-time (hence the relevant phase space) is non(multi)fractal. If this is not the case, some kind of 
extension appears to become necessary. Indeed, an everyday increasing list of physical anomalies are, 
here and there, being pointed out which defy (not to say that plainly violate!) the standard BG 
prescriptions. A {\it nonextensive} thermostatistics, which recovers the {\it extensive}, BG one as 
particular case, was proposed in 1988 \cite{tsallis,biblio}
which might correctly cover at least some of the known anomalies. Although a fair amount of what 
legitimately looks like being successful applications is nowadays accumulating, further verifications and 
deep understanding is needed and welcome. Computational work is highly desired since, on various 
grounds, the analytic discussion frankly appears to be untractable. Needless, of course, to say that more 
experimental and theoretical work is absolutely relevant to  exhibit the applicability and robustness of the 
ideas I intend to present herein.
In the present contribution, I propose some (hopefully relevant!) questions  that are right now open to 
such theoretical, experimental and computational contributions. 

Let us be more specific. As mentioned above, it is nowadays quite well known that a variety of physical 
systems exists
for which the powerful (and beautiful!) BG statistical
mechanics and standard thermodynamics present serious difficulties or
anomalies, which can occasionally achieve the status of just plain failures.
Within a long list that will be systematically focused on later on, we may mention at this point systems 
involving long-range interactions (e.g., $d=3$ gravitation)\cite{longrange}, long-range
microscopic memory (e.g., nonmarkovian stochastic processes, on which
much remains to be known, in fact)\cite{nonmarkovian,caceresbjp}, and, generally
speaking, conservative (e.g., Hamiltonian) or dissipative systems which in
one way or another involve a relevant space-time (hence, a relevant
phase space) which has a (multi)fractal-like structure. For instance,
pure-electron plasma two-dimensional turbulence\cite{huangdriscoll},
L\'evy anomalous diffusion\cite{shlesinger}, granular systems\cite{granular}, phonon-electron 
anomalous thermalization in ion-bombarded solids (\cite{koponen} and references therein), solar
neutrinos\cite{clayton}, peculiar velocities of galaxies\cite{galaxies},
inverse bremsstrahlung in plasma\cite{bremsstrahlung} and black
holes\cite{cosmology}, to cite a few, clearly appear to be (in some cases), or could possibly be (in 
others), concrete examples. The present status of these and others will be discussed in Sections III, IV 
and V.  

\section{FORMALISM}

\subsection{ENTROPY}

As an attempt to overcome at least some of these difficulties a proposal has been
advanced, one decade ago\cite{tsallis},
(see also \cite{curado,tsamepla}), which is based on a
generalized entropic form, namely
\begin{equation}\label{s}
S_q = k \frac{1-\sum_{i=1}^W p_i^q}{q-1}~~~~~~
\left(\sum_{i=1}^W p_i=1; \;q\in \cal{R} \right) \text{ ,}
\end{equation}
where $k$ is a positive constant and $W$ is the total number of
microscopic possibilities of the system (for the $q<0$ case, care
must be taken to exclude all those possibilities whose probability is not strictly
positive, otherwise $S_q$ would diverge; such care is not necessary
for $q>0$; due to this property, the entropy is said to be {\it expansible} for $q>0$).
 This expression recovers the usual BG entropy ($-k\sum_{i=1}^W p_i\;
\ln\;p_i$)  in the
limit  $q\rightarrow 1$.
The entropic index $q$ (intimately related to and determined by the
microscopic dynamics, as we shall mention later on) characterizes
the {\it degree of nonextensivity} reflected in the following
{\it pseudo-additivity}
entropy rule
\begin{eqnarray}
S_q(A+B)/k &=& [S_q(A)/k] +[S_q(B)/k]
\nonumber\\
 &+& (1-q)[S_q(A)/k][S_q(B)/k] \; ,
\end{eqnarray}
where $A$ and $B$ are two {\it independent} systems
 in the sense that the probabilities of $A+B$ {\em factorize} into
those of $A$ and of $B$ (i.e., $p_{ij}(A+B)=p_i(A)p_j(B)$). We immediately see that, since in all
cases $S_q \ge 0$ ({\it nonnegativity} property), $q<1,\;q=1$ and $q>1$ respectively correspond to
{\it superadditivity (superextensivity), additivity (extensivity) and
subadditivity (subextensivity)}. Eq. (2) exhibits a property which has apparently never been focused 
before, and which we shall from now on refer to as the {\it composability} property. It concerns the 
nontrivial fact that the entropy $S(A+B)$ of a system composed of two independent subsystems $A$ and 
$B$ can be calculated from the entropies $S(A)$ and $S(B)$ of the subsystems, {\it without any need of 
microscopic knowledge about $A$ and $B$, other than the knowledge of some generic universality class, 
herein the nonextensive universality class, represented by the entropic index $q$}, i.e., without any 
knowledge about the microscopic possibilities of $A$ and $B$ nor their associated probabilities. This 
property is so obvious for the BG entropic form that the (false!) idea that {\it all} entropic forms 
automatically satisfy it could easily install itself in the mind of most physicists. To show 
counterexamples, it is enough to check that the recently introduced Anteneodo-
Plastino's\cite{celiangelentropy} and Curado's\cite{curadobjp} entropic forms  satisfy a variety of 
interesting properties, and nevertheless are {\ not} composable.

The above pseudo-extensivity property can be equivalently written as follows:
\begin{equation}
\frac{\ln [1+(1-q)S_q(A+B)/k]}{1-q} = \frac{\ln [1+(1-q)S_q(A)/k]}{1-q} + \frac{\ln [1+(1-
q)S_q(B)/k]}{1-q}
\end{equation}
We come back onto this form later on in connection with Renyi's entropy.

Another important (since it eloquently exhibits the surprising effects
of nonextensivity) property is the following. Suppose that the set
of $W$ possibilities is arbitrarily separated into two subsets having
respectively $W_L$ and $W_M$ possibilities ($W_L+W_M=W$).
We define $p_L \equiv \sum_{i=1}^{W_L}p_i$ and $p_M \equiv
\sum_{i=W_L+1}^{W}p_i$, hence $p_L+p_M=1$. It can then be
straightforwardly established that
\begin{eqnarray}
S_q(\{p_i\})= &S_q(p_L,p_M)\; 
&+\;p_L^q\;S_q(\{p_i/p_L\})\;+\; p_M^q\;S_q(\{p_i/p_M\}) \; ,
\end{eqnarray}
where the sets $\{p_i/p_L\}$ and $\{p_i/p_M\}$ are the {\it conditional}
probabilities. This would precisely be the famous Shannon's property
{\it were it not for the fact that, in front of the entropies associated
with the conditional probabilities, appear $p_L^q$ and $p_M^q$ instead of
$p_L$ and $p_M$ }. This fact will play, as we shall see later on,
a central role in the whole generalization of thermostatistics.
Indeed, since the probabilities $\{p_i\}$ are generically numbers
between zero and unity, $p_i^q>p_i$ for $q<1$ and  $p_i^q<p_i$
for $q>1$, hence $q<1$ and $q>1$ will respectively privilegiate the
{\it rare} and the {\it frequent} events. This simple property lies at
the heart of the whole proposal. Santos has recently
shown\cite{robertojorge}, strictly following along the lines of
Shannon himself, that, if we assume (i) continuity (in the $\{p_i\}$) of the entropy,
(ii) increasing monotonicity of the entropy as a function of $W$ in
the case of equiprobability, (iii) property (2), and (iv) property (4),
then {\it only one entropic form exists, namely that given in
definition} (1). Of course, the generalization of Eq. (4) to the case where, instead of two, we have $R$ 
nonintersecting subsets ($W_1+W_2+...+W_R=W$) is straightforward\cite{tsallisvelho}. To be more 
specific, if we define
\begin{equation}
\pi_j \equiv \sum_{W_j\; terms}p_i\;\;\;\; (j=1,2,...,R)
\end{equation}
(hence $\sum_{j=1}^R \pi_j=1$),  Eq. (4) is generalized into
\begin{equation}
S_q(\{ p_i \})=S_q(\{\pi_j\})+ \sum_{j=1}^R \pi_j^q S_q(\{p_i/ \pi_j\})
\end{equation}
where we notice, in the last term, the emergence of what we shall soon introduce generically as the 
unnormalized $q$-expectation value (of the {\it conditional} entropies $ S_q(\{p_i/ \pi_j\})  $, in the 
present case).

Another interesting property is the following. The Boltzmann-Gibbs entropy $S_1$ satisfies the 
following relation:
\begin{equation}
-k\left[\frac{d}{d\alpha} \sum_{i=1}^W p_i^{\alpha}\right]_{\alpha=1} = 
-k\sum_{i=1}^Wp_i\;\ln\;p_i \equiv S_1 
\end{equation}
Moreover, Jackson introduced in 1909\cite{jackson} the following generalized differential operator
(applied to an arbitrary function $f(x)$):
\begin{equation}
D_q \;f(x) \equiv \frac{f(qx)-f(x)}{qx-x},
\end{equation}
which satisfies $D_1 \equiv lim_{q\rightarrow 1} D_q = \frac{d}{dx}$. Abe\cite{abe} recently
remarked that
\begin{equation}
-k\left[D_q \;\sum_{i=1}^W p_i^{\alpha}\right]_{\alpha=1} =k\frac{1-\sum_{i=1}^W p_i^q}{q-1} 
\equiv
S_q
\end{equation}
This property provides some insight into the generalized entropic form $S_q$ . Indeed, the
inspiration for its use in order to generalize the usual thermal statistics came\cite{tsallis} from
multifractals, and its applications concern, in one way or another, systems which
exhibit scale invariance. Therefore, its connection with Jackson's differential operator appears to
be rather natural. Indeed, this operator ``tests'' the function $f(x)$ under {\it dilatation} 
of $x$, in contrast to the usual derivative, which ``tests" it under {\it translation} of $x$.

Another property which no doubt must be mentioned in the present introduction  is that $S_q$ is 
consistent with Laplace's maximum ignorance principle, i.e., it is extremum at {\it equiprobability} ($p_i 
= 1/W\;\;\forall{i}$). This extremum is given by
\begin{equation}
S_q= k \frac{W^{1-q}-1}{1-q}\;\;\;\;\;\;(W \ge 1)
\end{equation}
which, in the limit $q \rightarrow 1$, reproduces Boltzmann's celebrated formula $S=k \ln\; W$ (carved 
on his marble grave in the Central Cemetery of Vienna). In the limit $W \rightarrow \infty$, $S_q$ 
diverges if $q \le 1$, and saturates at $k/(q-1)$ if $q>1$.

Finally, let us close the present set of properties by reminding that $S_q$ has, with regard to
$\{p_i\}$, a {\it definite concavity} for {\it all} values of $q$ ($S_q$ is always concave for $q>0$ and
always convex for $q<0$). In this sense, it contrasts with Renyi's entropy $S_q^R \equiv
(\ln \;\sum_{i=1}^W p_i^q)/(1-q) = \{\ln\;[1+(1-q)S_q/k]\}/(1-q)$, which does not have this property
for {\it all} values of $q$ .

Before addressing other relevant quantities, let us introduce the following convenient 
functions\cite{quimicanova}:
\begin{equation}
e_q^x \equiv [1+(1-q)\;x]^{1/(1-q)},\;\;\;\; \forall (x,q)
\end{equation}
(hence, $e_1^x = e^x$) with the definition supplement, for $q<1$, that $e_q^x =0$ if $1+(1-q)\;x \le 0$, 
(and analogously, for $q>1$, $e_q^x$ diverges at $x=1/(q-1)$) and
\begin{equation}
\ln_q\;x \equiv [x^{1-q}-1] / [1-q],\;\;\;\; \forall (x,q)
\end{equation}
(hence, $\ln_1\;x=\ln\;x$). We can verify easily that
\begin{equation}
e_q^{\;\ln_q\;x}=\ln_q\;e_q^x = x, \;\;\;\;\forall (x,q).
\end{equation}
For instance, Eq. (10) can be rewritten in the following Boltzmann-like form:
\begin{equation}
S_q = k\;\ln_q \;W
\end{equation}
Let us also introduce the following {\it unnormalized $q$-expectation value}:
\begin{equation}
\langle A \rangle_q \equiv \sum_{i=1}^W p_i^q \;A_i
\end{equation}
hence $\langle A \rangle_1$ corresponds to the standard mean value of a physical quantity $A$.

If our system is a generic quantum one, its probabilistic description is given by the density operator 
$\rho$, whose eigenvalues are the $\{p_i\}$. Then, the generalized entropy is given by
\begin{equation}
S_q=k\;\frac{1-Tr\;\rho^q}{q-1}\;\;\;\;\;(Tr\;\rho=1)
\end{equation}
and the unnormalized $q$-expectation value of an observable $A$ which does not necessarily commute 
with $\rho$ is given by
\begin{equation}
\langle A \rangle_q \equiv Tr\;\rho^qA
\end{equation}
Eq. (16) can be rewritten as follows:
\begin{equation}
S_q=-k\;\langle \ln_q\;\rho\rangle_q
\end{equation}
and also as
\begin{equation}
S_q=-k\;\langle \ln_{2-q}\;\rho\rangle_1
\end{equation}

If our system is a generic classical one, the relevant variables are typically continuous variables, and its 
probabilistic description is given by a distribution of probabilities $p({\bf r})$, where ${\bf r}$ is a 
dimensionless  variable in say a many-body phase space. Then, the generalized entropy is given by
\begin{equation}
S_q=k\;\frac{1-\int d{\bf r}\;[p({\bf r})]^q}{q-1}\;\;\;\;\;(\int d{\bf r}\;p({\bf r}) =1)
\end{equation}
and the unnormalized $q$-expectation value of an observable $A({\bf r})$ is given by
\begin{equation}
\langle A \rangle_q \equiv \int d{\bf r}\;[p({\bf r})]^q A({\bf r})
\end{equation}
Although we shall, in what follows, be illustrating the present formalism with the case of $W$ discrete 
microscopic possibilities, the generic quantum and classical discussions follow along the same lines, {\it 
mutatis mutandis}.

\subsection{CANONICAL ENSEMBLE}

Once we have a generalized entropic form, say that given in Eq. (1) (or an even more general one, or a 
different one), we can use it in a variety of ways. For instance, if we are interested in information theory, 
some optimization algorithms, image processing, among others, we can take advantage of a particular 
form in different ways. See, for instance, \cite{curadobjp,tsallisvelho,generalentropies,galore} and 
references therein, where it can be verified that not less than 25 (!) different entropic forms have 
received, along the years, a great variety of technological and mathematical applications. For instance, 
the Renyi entropy mentioned above has been quite useful in the geometrical characterization of strange 
attractors and similar multifractal structures (see \cite{procaccia} and references therein). 

However, if our primary interest is Physics, this is to say the (qualitative and quantitative) description and 
possible understanding of phenomena occurring in Nature, then we are naturally led to use the available 
generalized entropy in order to generalize statistical mechanics itself and, if unavoidable, even 
thermodynamics. It is along this line that we shall proceed from now on (see also \cite{plastinosbjp}). To 
do so, the first nontrivial (and quite ubiquitous) physical situation is that in which a given system is in 
contact with a thermostat at temperature $T$. To study this, we shall follow along Gibbs' path and focus 
the so called {\it canonical ensemble}. More precisely, to obtain the thermal equilibrium distribution  
associated with a conservative (Hamiltonian) physical system in contact with the thermostat we shall 
extremize $S_q$ under appropriate constraints. These constraints are\cite{tsamepla} 
\begin{equation}
\sum_{i=1}^W p_i=1\;\;\;\;\;\;\;\;\;\;\;\;\;\;\;\;\;\;\;\;\;\;\;(norm\; constraint)
\end{equation}
and
\begin{equation}
\langle \langle \epsilon_i \rangle \rangle_q \equiv \frac{\sum_{i=1}^W p_i^q \epsilon_i }{ 
\sum_{i=1}^W p_i^q }= U_q\;\;\;(energy \;constraint)
\end{equation}
where $\{\epsilon_i\}$ are the eigenvalues of the Hamiltonian of the system. We shall refer to 
$\langle \langle...\rangle \rangle_q$ as the {\it normalized q-expectation value} and to $U_q$ as the {\it 
generalized internal energy} (assumed {\it finite} and fixed). It is clear that, in the $q \rightarrow 1$ 
limit, these quantities recover the standard mean value and internal energy respectively. We immediately 
verify that, for any observable,
\begin{equation}
\langle \langle...\rangle \rangle_q=\frac{\langle ...\rangle_q}{\langle 1\rangle_q}
\end{equation}

The outcome of this optimization procedure is given by
\begin{eqnarray}
p_i &=& \frac{ \left[1-(1-q) \beta (\epsilon_i-
U_q)/\sum_{j=1}^W \left(p_j\right)^q
\right]^{\frac{1}{1-q}}}{\bar{Z}_q}
\end{eqnarray}
with
\begin{eqnarray} \label{zq}
\bar{Z}_q(\beta)&\equiv&\sum_{i=1}^W\left[1-(1-q) \beta (\epsilon_i- U_q)
/{ \sum_{j=1}^W \left(p_j\right)^q} \right]^{\frac{1}{1-q}} 
\end{eqnarray}
It can be shown that, for the case $q<1$, the expression of the equilibrium distribution is complemented 
by the auxiliary condition that $p_i = 0$ whenever the argument of the function becomes negative ({\it 
cut-off} condition).
Also, it can be shown\cite{tsamepla} that
\begin{equation}
1/T= \partial S_q /\partial U_q,\;\;\forall q\;\;\;(T \equiv 1/(k \beta)).
\end{equation}
Furthermore, it is important to notice that, if we add a constant $\epsilon_0$ to all $\{\epsilon_i\}$, we 
have (as it can be self-consistently proved) that
$U_q$ becomes $U_q+ \epsilon_0$, which leaves invariant
the differences $\{\epsilon_i-U_q\}$, which, in turn,
(self-consistently) leaves {\it invariant} the set of probabilities
$\{p_i\}$, hence {\it all} the thermostatistical quantities.
It is also trivial to show that, for the independent systems $A$ and $B$ mentioned previously,
$U_q(A+B) = U_q(A) + U_q(B)$,
thus recovering the same form of the standard ($q=1$) thermodynamics. 

It can be shown that the following relations hold:
\begin{equation}
\sum_{i=1}^W (p_i)^q=(\bar{Z}_q)^{1-q},
\end{equation}
\begin{equation}
F_q \equiv U_q - T S_q = -\frac{1}{\beta} \frac{(Z_q)^{1-q}-1}{1-q}
\end{equation}
and
\begin{equation}
U_q = -\frac{\partial}{\partial \beta}\frac{(Z_q)^{1-q}-1}{1-q},
\end{equation}
where
\begin{equation}
\frac{(Z_q)^{1-q}-1}{1-q}=\frac{(\bar{Z}_q)^{1-q}-1}{1-q}-\beta U_q.
\end{equation}

Let us now make an important remark. If we take out as factors,
 in both numerator and denominator of Eq. (25),
the quantity $ \left[1+(1-q)\beta U_q/{\sum_{j=1}^W
\left(p_j\right)^q }\right]$ ,
and then cancel them, we obtain
\begin{equation} \label{pp}
p_i(\beta)=\frac{[1-(1-q)\beta^{\prime} \epsilon_i]^{\frac{1}{1-q}}}{Z_q^{\prime}}  
\;\;\;\;\;\;
\left(Z_q^{\prime} \equiv \sum_{j=1}^W [1-(1-q)\beta^{\prime}
\epsilon_j]^{\frac{1}{1-q}}\right)  
\end{equation}
with
\begin{equation} \label{bp}
\beta^{\prime}= \frac{\beta}{\sum_{j=1}^W \left(p_j\right)^q
+\;(1-q)\beta U_q}\;\;\;\;\;( T^{\prime} \equiv 1/(k {\beta}^{\prime}))
\end{equation}
where $\beta^{\prime}$ is an increasing function of $\beta$ \cite{adriano}.

Let us now address the all important question of the connection between experimental numbers (those 
provided by measurements), and the quantities that appear in the theory. The definition of $U_q$ 
suggests the following {\it normalized}
$q$-{\it expectation values}
\begin{equation}\label{oq}
O_q \equiv \langle \langle  O_i  \rangle \rangle_q \equiv
\frac{\sum_{i=1}^W p_i^q O_i}{ \sum_{i=1}^W p_i^q} 
\end{equation}
where $O$ is any observable which commutes with the Hamiltonian, hence with $\rho$. If it does not 
commute, Eq. (34) is generalized into
\begin{equation}
O_q \equiv \frac{Tr \rho^q\;O}{Tr \rho^q}
\end{equation}
Consistently, $O_q$ is the mathematical object to be identified with the numerical value provided by the 
experimental measure. Later on, we come back onto this crucial point.

At this point let us make some observations about the set of {\it escort probabilities}\cite{escort} 
$\{P_i^{(q)}\}$
defined through
\begin{equation}
P_i^{(q)} \equiv \frac{p_i^q}{\sum_{j=1}^Wp_j^q}\;\;\;\;
(\sum_{i=1}^W P_i^{(q)} = 1)
\end{equation}
from which follows the dual relation
\begin{equation}
p_i = \frac{[P_i^{(q)}]^{\frac{1}{q}}}{\sum_{j=1}^W [P_j^{(q)}]^{\frac{1}{q}}}.
\end{equation}
The $W=2$ illustration of $P_i^{(q)}$ is shown in Fig. 1. As anticipated, $q<1$ ($q>1$) privileges the 
{\it rare} ({\it frequent}) events.

Let us first comment that Eqs. (36) and (37) have, within the present formalism, a role somehow 
analogous to the direct and inverse Lorentz transformations in Special Relativity (see \cite{tsallisabe} 
and references therein). Second, we notice that $O_q$ becomes an usual mean value
when expressed in terms of the probabilities $\{P_{i}^{(q)}\}$, i. e.,
\begin{equation}
O_q \equiv \frac{\sum_{i=1}^W  p_i^q O_i}{\sum_{j=1}^W p_j^q} =
\sum_{i=1}^W P_i^{(q)} O_i \; .
\end{equation}
and
\begin{equation}
\sum_{i=1}^W P_i^{(q)} \epsilon_i = U_q \;.
\end{equation}
The final equilibrium distribution reads
\begin{equation}
P_i^{(q)} = \frac{[1-(1-q)\beta'
\epsilon_i]^{\frac{q}{1-q}}}{\sum_{k=1}^W{[1-(1-q)\beta'\epsilon_k]^{\frac{q}{1-q}}}}\;.
\end{equation}
If the energy spectrum $\{\epsilon_i\}$ is associated with the set of {\it degeneracies} $\{g_i\}$, then the 
above probability leads to the following one (associated with the {\it level} $\epsilon_i$ and not the {\it 
state} $i$)
\begin{equation}
P(\epsilon_i) = \frac{g_i[1-(1-q)\beta'
\epsilon_i]^{\frac{q}{1-q}}}{\sum_{all \;levels}g_k{[1-(1-q)\beta'\epsilon_k]^{\frac{q}{1-q}}}}
\end{equation}
If the energy spectrum $\{\epsilon_i\}$ is so dense that can practically be considered as a continuum, 
then the discrete degeneracies yield the function {\it density of states} $g(\epsilon)$, hence
\begin{equation}
P(\epsilon) = g(\epsilon) \frac{[1-(1-q)\beta'
\epsilon]^{\frac{q}{1-q}}}{\int d\epsilon^{\prime}\; g(\epsilon^{\prime}){[1-(1-
q)\beta'\epsilon^{\prime}]^{\frac{q}{1-q}}}}
\end{equation}
The density of states is of course to be calculated for every specific Hamiltonian (given the boundary 
conditions). For instance, for a $d$-dimensional ideal gas of particles or quasiparticles, it is 
given\cite{conceicao} by $
g(\epsilon) \propto \epsilon^{\frac{d}{\delta}-1}$, where $\delta$ is the exponent characterizing the 
energy spectrum $\epsilon \propto  K^{\delta}$ where $K$ is the wavevector (e.g., $\delta=1$ 
corresponds to the harmonic oscillator, $\delta=2$ corresponds to a nonrelativistic particle in an infinitely 
high square well, etc).
In Figs. 2 and 3 we see typical energy distributions for the particular case of a constant density of states. 
Of course, the $q=1$ case reproduces the celebrated Boltzmann factor. Notice the cut-off for $q<1$ and 
the long algebraic tail for $q>1$.

All the above considerations refer, strictly speaking, to {\it thermodynamic equilibrium}. The word {\it 
thermodynamic} makes allusion to ``very large" ($N \rightarrow \infty$, where $N$ is the number of 
microscopic particles of the physical system). The word {\it equilibrium} makes allusion to 
asymptotically large times ($t\rightarrow \infty$ limit) (assuming a stationary state is eventually 
achieved). The question arises: {\it which of them first?} Indeed, although both possibilities clearly 
deserve the denomination "thermodynamic equiulibrium", nonuniform convergences might be involved 
in such a way that $lim_{N \rightarrow \infty}\; lim_{t \rightarrow \infty}$ could differ from $lim_{t 
\rightarrow \infty}\; lim_{N \rightarrow \infty}$. To illustrate this situation, let us imagine a classical 
Hamiltonian system including two-body interactions decaying at long distances as $1/r^{\alpha}$ in a 
$d$-dimensional space, with $\alpha \ge 0$. If $\alpha > d$ the interactions are essentially {\it short}-
ranged, the two limits just mentioned are basically interchangeable, and the prescriptions of standard 
statistical mechanics and thermodynamics are valid, thus yielding {\it finite} values for all the physically 
relevant quantities. In particular, the Boltzmann factor certainly describes reality, as very well known. 
But, if $0 \le \alpha \le d$, nonextensivity is expected to emerge, the order of the above limits becomes 
important because of nonuniform convergences, and the situation is certainly expected to be more subtle. 
More precisely, a crossover (between $q \ne 1$ and $q=1$ behaviors) is expected to occur at $t=\tau(N)$. 
If $lim_{N \rightarrow \infty}  \tau(N)= \infty$, then we would indeed have {\it two (or even more) 
different and equally legitimate states of thermodynamic equilibrium}, instead of the familiar unique 
state. The conjecture is illustrated in Fig. 4.

A wealth of works has shown that the above described nonextensive statistical mechanics retains much
of the formal structure of the standard theory. Indeed, many important
properties have been shown to be $q$-{\it invariant}. Among them, it is mandatory to mention \\
(i) the Legendre transformations structure of thermodynamics \cite{curado,tsamepla};\\
(ii) the H-theorem (macroscopic time irreversibility), more precisely, that, in the presence of some 
irreversible physical evolution, $dS_q/dt \ge 0,\;=0$ and $\le 0$ if $q>0,\;=0$ and $<0$, respectively, the 
equalities holding for equilibrium\cite{ananias,boplatsa};\\
(iii)  the Ehrenfest theorem (correspondence principle between classical and quantum 
mechanics)\cite{ehrenfest};\\
(iv)  the Onsager reciprocity theorem (microscopic time reversibility)\cite{onsager,rajagopal};\\
(v)   the Kramers and Wannier relations (causality)\cite{rajagopal};\\
(vi)  the factorization of the likelihood function (Einstein' 1910 reversal of Boltzmann's 
formula)\cite{tsallisvelho};\\
(vii)  the Bogolyubov inequality\cite{bogoliubov};\\
(viii)  thermodynamic stability (i.e., a definite sign for the specific heat)\cite{stability};\\
(ix)  the Pesin equality\cite{pesin}.

In contrast with the above quantities and properties, which are $q$-invariant, some others {\it do} depend 
on $q$, such as\\
(i)  the specific heat \cite{specheat};\\
(ii)  the magnetic susceptibility \cite{fluctuationdissipation};\\
(iii)  the fluctuation-dissipation theorem (of which the two previous properties can be considered as 
particular cases) \cite{fluctuationdissipation};\\
(iv)  the Chapman-Enskog expansion, the Navier-Stokes equations and related transport 
coefficients\cite{boghosianbjp};\\
(v)  the Vlasov equation\cite{plastinos,boghosian};\\
(vi)  the Langevin, Fokker-Planck and Lindblad 
equations\cite{langevin,correlatedplastinos,bukman,correlatedvarious,lindblad};\\
(vii)  stochastic resonance\cite{wiobjp};\\
(viii)  the mutual information or Kullback-Leibler entropy\cite{boplatsa,tsallismutual}.

A remark is necessary with regard to both sets just mentioned. Indeed, these properties have in fact been 
studied, whenever applicable, within unnormalized $q$-expectation values for the constraints, rather than 
within the normalized ones that we are using herein. Nevertheless, they still hold because they have been 
established for fixed $\beta$, which, through Eq. (33), implies fixed $\beta^{\prime}$.

Finally, let us mention various important theoretical tools which enable the thermostatistical discussion 
of complex nonextensive systems, and which are now available (within the unnormalized and/or 
normalized versions for the $q$-expectation values) for arbitrary $q$. We refer to \\
(i)  Linear response theory\cite{rajagopal};\\
(ii)  Perturbation expansion\cite{lemame};\\
(iii)  Variational method (based on the Bogoliubov inequality)\cite{lemame};\\
(iv)  Many-body Green functions\cite{ramele};\\
(v)  Path integral and Bloch equation\cite{lemame2}, as well as related properties\cite{rajagopalbjp};\\
(vi)  Quantum statistics and those associated with the Gentile and the Haldane exclusion 
statistics\cite{haldane,ugurquantum,gentile};\\
(vii)  Simulated annealing and related optimization, Monte Carlo and Molecular dynamics 
techniques\cite{tsasta,penna,straub,schulte,okamoto,serra,xiang,lemes,kleber,muntsa,nishimori}.\\

\section{THEORETICAL EVIDENCES AND CONNECTIONS}

\subsection{Levy-type anomalous diffusion}

An enormous amount of phenomena in Nature follow the Gaussian distribution: measurement error 
distributions, height and weight distributions in biological individuals of given species, Brownian motion 
of particles in fluids, Maxwell-Boltzmann distribution of particle velocities in a variety of systems, noise 
distribution in uncountable electronic devices, energy fluctuations at thermal equilibrium of many 
systems, to only mention a few. Why is it so? Or, equivalently, what is their (thermo)statistical 
foundation? This fundamental problem has already been addressed, particularly by Montroll, and 
satisfactorily answered (see \cite{shlesinger} and references therein). The answer basically relies onto 
two pillars, namely the BG entropy and the standard central limit theorem. However, the Gaussian is {\it 
not} the only ubiquitous distribution: we also similarly observe Levy distributions (in 
micelles\cite{micelles}, supercooled laser\cite{laser}, fluid motion\cite{fluid}, wandering 
albatrosses\cite{albatroz}, heart beating\cite{heart}, financial 
data\cite{mandelbrot,mandelbrot2,financialstanley}, among many others). So, once again, what is the 
(thermo)statistical foundation of their ubiquity? This relevant question has also been addressed, once 
again by Montroll and collaborators\cite{shlesinger} among others. In this case however, a satisfactory 
answer has been missing for  a long time. The first successful step toward (what we believe to be) the 
solution was performed in 1994 by Alemany and Zanette\cite{alza}, who showed that the generalized 
entropic form $S_q$ was able to provide a {\it power}-law (instead of the {\it exponential}-law 
associated with Gaussians) decrease at long distances. Many other works followed along the same 
lines\cite{various,tsalesoma}. In \cite{tsalesoma} it was exhibited how the Levy-Gnedenko central limit 
theorem also plays a crucial role by transforming, through successive iterations of the jumps, the power-
law obtained from optimization of $S_q$ into the specific power-law appearing in Levy distributions. 
Summarizing, in complete analogy with the above mentioned Gaussian case  (and which is recovered in 
the more powerful present formalism as the $q=1$ particular case), the answer once again relies onto two 
pillars, which now are the generalized entropy $S_q$ and the Levy-Gnedenko central limit theorem.

The arguments have been very recently re-worked\cite{pratsa} out on the basis of the normalized $q$-
expectation values introduced in \cite{tsamepla}. These are the results that we briefly recall here.

Let us write $S_q$ as follows:
\begin{equation}
S_q[p(x)]=k\frac{1-\int_{-\infty}^{\infty} \frac{dx}{\sigma}\;[\sigma\;p(x)]^q}{q-1}
\end{equation}
where $x$ is the distance of one jump, and $\sigma>0$ is the characteristic length of the problem.
We optimize (maximize if $q>0$, and minimize if $q<0$) $S_q$ with the norm constraint $\int_{-
\infty}^{\infty} dx\; p(x) =1$, as well as with the constraint
\begin{equation}
\langle \langle x^2 \rangle \rangle_q \equiv \frac{\int_{-\infty}^{\infty} dx\;x^2\;[p(x)]^q}{\int_{-
\infty}^{\infty} dx\;[p(x)]^q} = \sigma^2
\end{equation}
We straightforwardly obtain the following one-jump distribution.\\
If $q>1$:\\
\begin{equation}
p_q(x)= \frac{1}{\sigma} \Bigl[ \frac{q-1}{\pi\;(3-q)} \Bigr]^{1/2} \frac{\Gamma(\frac{1}{q-
1})}{\Gamma(\frac{3-q}{2(q-1)})} \frac{1}{\bigl[1+\frac{q-1}{3-q}\;\frac{x^2}{\sigma^2}\bigr]^{1/(q-
1)}}
\end{equation}
If $q=1$:\\
\begin{equation}
p_q(x)=\frac{1}{\sigma}\Bigl[\frac{1}{2\pi}\Bigr]^{1/2}\; e^{-(x/\sigma)^2/2}
\end{equation}
If $q<1$:\\
\begin{equation}
p_q(x)= \frac{1}{\sigma} \Bigl[ \frac{1-q}{\pi\;(3-q)} \Bigr]^{1/2}
\frac{\Gamma(\frac{5-3q}{2(1-q)})}{\Gamma(\frac{2-q}{1-q})} 
\Bigl[1-\frac{1-q}{3-q}\;\frac{x^2}{\sigma^2}\Bigr]^{1/(1-q)}
\end{equation}
if $|x| <\sigma [(3-q)/(1-q)]^{1/2}$ and zero otherwise.

We see that the support of $p_q(x)$ is compact if $q \in (-\infty,1)$, an exponential behavior is obtained 
if $q=1$, and a power-law tail is obtained if $q>1$ (with $p_q(x) \propto (\sigma/x)^{2/(q-1)}$ in the 
limit $|x|/\sigma \rightarrow \infty$). Also, we can check that $\langle \langle x^2 \rangle \rangle_1= 
\langle x^2 \rangle_1=\int_{-\infty}^{\infty} dx\;x^2\;p_q(x)$ is finite if $q<5/3$ and diverges if $5/3 \le 
q \le 3$ (the norm constraint cannot be satisfied if $q \ge 3$). Finally, let us mention that the Gaussian 
($q=1$) solution is recovered in both limits $q \rightarrow 1+0$ and $q \rightarrow 1-0$ by using  the 
$q>1$ and the $q<1$ solutions respectively. This family of solutions is illustrated in Fig. 5.

We focus now the $N$-jump distribution $p_q(x,N)=p_q(x)*p_q(x)*...*p_q(x)$ ($N$-folded convolution 
product). If $q<5/3$, the standard central limit theorem applies, hence, in the limit $N \rightarrow 
\infty$, we have
\begin{equation}
p_q(x,N) \sim \frac{1}{\sigma} \Bigl[\frac{5-3q}{2\pi (3-q)N}\Bigr]^{1/2} \exp{\Bigl(-\frac{ 5-
3q}{2(3-q)N} \frac{x^2}{  \sigma^2}}\Bigr) 
\end{equation}
i.e., the attractor in the distribution space is a Gaussian, consequently we have {\it normal} diffusion. If, 
however, $q>5/3$, then what applies is the Levy-Gnedenko central limit theorem, hence, in the limit $N 
\rightarrow \infty$, we have
\begin{equation}
p_q(N,x) \sim L_{\gamma}(x/N^{1/\gamma})
\end{equation}
where $L_{\gamma}$ is the Levy distribution with index $\gamma <2$ given by
\begin{equation}
\gamma=\frac{3-q}{q-1}\;\;\;\;\;(5/3<q<3)
\end{equation}
Through the Fourier transforms of both Eq. (48) and (49), we can characterize the {\it width} $\Delta_q$ 
(dimensionless diffusion coeffiecient) of $p_q(x,N)$. We obtain
\begin{equation}
 \Delta_q \equiv \frac{3-q}{5-3q}   \;\;\;\;\;\;\;(q<5/3)
\end{equation}
and
\begin{equation}
\Delta_q=\frac{2}{\pi^{1/2}}\Bigl[ \frac{q-1}{3-q}\Bigr]^{\frac{3-q}{2(q-1)}}\; 
\Gamma\Bigl[\frac{3q-5}{2(q-1)}\Bigr]\;\;\;\;\;\;(5/3<q<3)
\end{equation}
These results are depicted in Fig. 6. This result should be measurable in specifically devised experiments. 
More details can be found in \cite{pratsa} and references therein. What we wish to retain in this short 
review is that the present formalism is capable of (thermo)statistically founding, in an unified and simple 
manner, both Gaussian and Levy behaviors, very ubiquitous in Nature (respectively associated with 
normal diffusion and a certain type of anomalous superdiffusion).

\subsection{Correlated-type anomalous diffusion}

There are some phenomena exhibiting anomalous (super and sub) diffusion of a type which differs from 
the one discussed in the previous subsection. We refer to the so called {\it correlated}-type of diffusion. 
We consider here a quite large class of them, namely those associated with the following generalized, 
Fokker-Planck-like equation:
\begin{equation}
\frac{\partial}{\partial t}[p(x,t)]^{\mu}  =-\frac{\partial}{\partial x}\{F(x)[p(x,t)]^{\mu}\} + 
D\frac{\partial^2}{x^2}[p(x,t)]^{\nu}
\end{equation}
where $(\mu,\nu) \in \cal{R}$$^2$, $D$ is a dimensionless diffusion-like constant, $F(x) \equiv -dV/dx$ 
is a dimensionless external force (drift) associated with a potential $V(x)$, and $(x,t)$ is a dimensionless 
$1+1$ space-time. If $\mu =1$, we can interpret $p(x,t)$ as a probability distribution since $\int 
dx\;p(x,t)=1,\;\; \forall t$ can be satisfied. If $\mu \ne 1$, then $p(x,t)$ must be seen as  a  density 
function. The word ``correlated" is frequently used in this context due to the fact that 
$D(\partial^2/\partial x^2)[p(x,t)]^{\nu} = (\partial / \partial x)\{D\nu [p(x,t)]^{\nu-1}\;(\partial / \partial 
x)\;p(x,t)\}$, i.e., an effective  diffusion emerges, for $\nu \ne 1$, which depends on $p(x,t)$ itself, a 
feature which is natural in the presence of correlations. The $\mu =1$  particular case of this nonlinear 
equation is commonly denominated ``Porous medium equation", and corresponds to a variety of physical 
situations (see \cite{bukman} and references therein for several examples).

The first connection of Eq. (53) with the present nonextensive statistical mechanics was established in 
1995 by Plastino and Plastino\cite{correlatedplastinos}. They considered a particular case, namely $\mu 
=1$ and $F(x)=- k_2 x$ with $k_2>0$ (so called Uhlenbeck-Ornstein processes), and found an exact 
solution which has the form of Eq. (43-45). Their work was generalized in \cite{bukman} where arbitrary 
$\mu $ and $F(x)=k_1-k_2x$ were considered. The explicit exact solution of Eq. (53), for {\it all values} 
of $(x,t)$, was once again found by proposing an Ansatz of the form of Eqs. (45-47), i.e., the form which 
optimizes $S_q$ with the associated simple constraints. This form eventually turns out to be the 
Barenblatt one, useful in related problems. Here, let us restrict ourselves to just reproduce the exact 
solution of Eq. (53) assuming that $p(x,0)=\delta(x)$, this is to say, a Dirac delta distribution. We 
obtain\cite{bukman}
\begin{equation}
p_q(x,t)= \frac{\{1-(1-q) \beta(t) [x-x_M(t)]^2\}^{1/(1-q)}}{Z_q(t)}
\end{equation}
where
\begin{equation}
q=1+\mu -\nu  
\end{equation}
and
\begin{equation}
\frac{\beta(t)}{\beta(0)}=\Bigl[\frac{Z_q(0)}{Z_q(t)}\Bigr]^{2\mu } 
\end{equation}
with
\begin{equation}
Z_q(t)=Z_q(0) \Bigl[ \Bigl(1-\frac{1}{K_2} \Bigr)e^{-t/ \tau} + \frac{1}{K_2}\Bigr]^{1/(\mu +\nu)  },
\end{equation}
\begin{equation}
K_2 \equiv \frac{k_2}{2\nu D \beta(0)[Z_q(0)]^{\mu -\nu  }}
\end{equation}
and
\begin{equation}
\tau \equiv \frac{\mu  }{k_2(\mu + \nu) }
\end{equation}

Summarizing, by using the form which optimizes $S_q$, it has been possible to find the physically 
relevant solution of a {\it nonlinear} equation in partial derivatives with {\it integer} derivatives. It can 
be shown\cite{alaor} that the problem that was solved in the previous subsection corresponds to a {\it 
linear} equation in partial derivatives but with {\it fractional} derivatives. We believe that we are 
allowed to say that an unusual mathematical versatility has been observed, within the present 
nonextensive formalism, in this couple of nontrivial examples of anomalous diffusion.

\subsection{Stellar polytropes and other self-gravitating systems}

The present formalism has been applied to a variety of astrophysical\cite{plastinos,boghosian} and 
cosmological\cite{cosmos} self-gravitating systems.  In some sense, this is something natural to do given 
the long range of the gravitational interaction. This was, in fact, the first physical application of 
nonextensive statistics. We do not intend here to reproduce details. Our present aim is to remind that it is 
well known in astrophysics that, within the standard thermodynamical approaches, it is not possible to 
{\it simultaneously} have {\it finite} values for the total energy, entropy and mass of a self-gravitating 
system. Plastino and Plastino were the first to show, in 1993, that this physically desirable situation {\it 
can} be achieved if we allow $q$ to sufficiently differ from unity ! In fact, it can be shown (by 
considering the Vlasov equation in $D$-dimensional Schuster spheres) that the problem becomes a 
mathematically well posed one if $q<q^*$, where the critical value $q^*$ is given by
\begin{equation}
q^* = \frac{8-(D-2)^2}{8-(D-2)^2+2(D-2)}.
\end{equation}
For $D=3$ we recover the 7/9  relatively known value. Also, we notice that $D=2$ implies $q^*=1$, 
which is very satisfactory since it is known that $D<2$ gravitation is tractable within standard 
thermodynamics.

\subsection{Zipf-Mandelbrot law}

The problem we focus here first appeared in Linguistics. However, its relevance is quite broad, as it will 
soon become clear. Suppose we take a given text, say Cervante's {\it Don Quijote}, and order all of its 
words from the most to the less frequent; we refer to the ordered position of a given word as its {\it rank} 
$R$ (low rank means high frequency $\omega$ of appearance in the text, and high rank means low 
frequency). Zipf\cite{zipf} discovered that, in this as well as in a variety of similar problems, the 
following law is satisfied:
\begin{equation}
\omega = A\;R^{-\xi}   \;\;\;\;\;\; (Zipf \;law)
\end{equation}
where $A>0$ and $\xi >0$ are constants. Later on, Mandelbrot\cite{mandelbrot} suggested that such 
behavior was reflecting a kind of fractality hidden in the problem; moreover, he suggested how the Zipf 
law could be numerically improved:
\begin{equation}
\omega=\frac{A}{(D+R)^{\xi}}\;\;\;\;\;(Zipf-Mandelbrot\;law)
\end{equation}
This expression has been useful in a variety of analysis. The connection  we wish to mention here is that 
in 1997 Denisov\cite{denisov} showed that, by extending (to arbitrary $q$) the well known Sinai-Bowen-
Ruelle thermodynamical formalism of symbolic dynamics (i.e., by considering $S_q$ instead of $S_1$), 
the Zipf-Mandelbrot law can be {\it deduced}.  He obtained
\begin{equation}
\xi=\frac{1}{q-1}
\end{equation}
and $D=d/(q-1)$ ($d$ being a positive constant), i.e.,
\begin{equation}
\omega \propto \frac{1}{[1+(q-1)R/d]^{1/(q-1)}}\;\;\;\;(q>1)
\end{equation}
Clearly, to make the discussion complete, a model would be welcome, which would provide quantities 
such as $q$ and $d$. Nevertheless, Denisov's arguments have the deep interest of explicitly exhibiting 
that the Zipf-Mandelbrot law can be seen as having a  nonextensive foundation. Fittings with 
experimental data will be shown later on in connection with the citations of scientific papers.

\subsection{Theory of financial decisions; Risk aversion}

An important problem in the theory of financial decisions is how to take into account extremely relevant 
phenomena such as the risk aversion human beings (hence financial operators) quite frequently feel. This 
kind of problem has, since long, been extensively studied by Tversky\cite{tversky} and co-workers. The 
situation can be illustrated as follows. What do you prefer, to earn 85,000 dollars or to play a game in 
which you have 0.15 probability of earning nothing and 0.85 probability of earning 100,000 dollars ? In 
fact, most people  prefer take the money. The problem of course is the fact that the expectation value for 
the gain is one and the same (more precisely 85,000 dollars) for both choices, and therefore this 
mathematical tool does not reflect reality ! The same problem appears if one expects to loose 85,000 and 
the chance is given for playing a game in which, if you win, you pay nothing, but, if you loose, you pay 
100,000 dollars. In this case, most people choose to play. So, the experimental facts are that most human 
beings are {\it risk-averse} when they expect to gain, and {\it risk-seeking} when they expect to loose ! 
The problem is how to put this into mathematics.

Following \cite{tsallisrisk}, let us introduce, for the above gain problem, normalized $q$-expectation 
values as follows:
\begin{equation}
\langle \langle gain \rangle \rangle_q^{take\; the\; money} =85,000
\end{equation}
and
\begin{equation}
\langle \langle gain \rangle \rangle_q^{play\; the\; game} =\frac{100,000 \times 0.85^q+0 \times 
0.15^q}{0.85^q+0.15^q}=\frac{100,000 \times 0.85^q}{0.85^q+0.15^q}
\end{equation}
Since most people would prefer the money, this means that most people have $q<1$ for this particular 
decision problem. 

For the loss probem we have:
\begin{equation}
\langle \langle gain \rangle \rangle_q^{take\; the\; money} =-85,000
\end{equation}
and
\begin{equation}
\langle \langle gain \rangle \rangle_q^{play\; the\; game} =\frac{-100,000 \times 0.85^q+0 \times 
0.15^q}{0.85^q+0.15^q}=\frac{-100,000 \times 0.85^q}{0.85^q+0.15^q}
\end{equation}
Since in this case most people would prefer to play, this means that, consistently with the previous result,
 most people have $q<1$ for the particular decision problem we are considering now. In some sense, we 
have some epistemological progress ! Indeed, the statement ``most people have (for this type of amount 
of money) $q<1$", {\it unifies} the previous  two {\it separate} statements concerning expectation to gain 
and expectation to loose. 

Let us address now the following question: how can we measure the value of $q$ associated with a
particular individual ? We illustrate this interesting point with the example of the gain. The person is 
asked to choose between having $V$ dollars or playing a game in which, if the person wins, the prize 
will be $100,000$ dollars and, if the person looses, he (she) will receive nothing. As before, the person is 
informed that his (her) probability of winning is 0.85 (hence, the probability of loosing is 0.15). Then we
keep gradually changing the value $V$ and asking what is the preference. At a certain critical value, 
noted $V_c$, the person will change his (her) mind. Then, the value of $q$ to be associated with that 
person, for that problem, is given by the following equality
\begin{equation}
\frac{100,000 \times 0.85^q}{0.85^q+0.15^q}=V_c
\end{equation}
(See Fig. 7). The ideally rational operator corresponds to $q=1$. For this gain problem, the risk-averse 
operators correspond to $q<1$, and the risk-seeking ones to $q>1$.
It is clear that models for stock exchange can be formulated by using these remarks. Such an effort is 
presently in progress\cite{alexfinance}.

\subsection{Physiology of vision}

Physiological perceptions such as the visual perception are since long known to focus upon {\it rare} 
events (e.g., a red spot on a white wall). Barlow\cite{barlow}, among others, has recurrently stressed our 
attention on the fact that, at the action decision level, the various possibilities should enter with a weight 
proportional to $-\ln p_i$, and {\it not} proportional to $p_i$, $p_i$ being the a priori probability of 
occurrence  of that particular event; indeed, $-\ln p_i$ diverges when $p_i \rightarrow 0$. He even argues 
that evolutionary arguments hold very well together with such hypothesis. To privilege rare events is 
precisely what happens, in the present formalism, whenever $q<1$. Let us be more specific: if we 
consider the $0<q<<1$ limit, we obtain\cite{tsallisvision}
\begin{equation}
S_q/k = \frac{1-\sum_{i=1}^W p_i^q}{q-1} \sim W-1 +q[W-1-\sum_{i=1}^W (-\ln p_i)],
\end{equation}
\begin{equation}
\langle O \rangle_q \equiv \sum_{i=1}^Wp_i^q\;O_i \sim \sum_{i=1}^W O_i - q \sum_{i=1}^W 
(-\ln p_i)\;O_i
\end{equation}
and
\begin{equation}
\langle \langle O \rangle \rangle_q \equiv \frac{\sum_{i=1}^Wp_i^q\;O_i}{\sum_{i=1}^Wp_i^q } \sim 
\frac{\sum_{i=1}^W O_i}{W} \Bigl\{1+q\Bigl[\frac{\sum_{i=1}^W (-\ln p_i)}{W} - 
\frac{\sum_{i=1}^W (-\ln p_i)\;O_i}{\sum_{i=1}^W O_i}     \Bigr] \Bigr\}
\end{equation}
where $O$ is an arbitrary observable. Leaving aside several constant quantities that appear above, we 
immediately observe the prominent role which $-\ln p_i$ plays in these expressions. Consistently, the $q 
\rightarrow 0$ limit of the present formalism could well be of some utility in the theoretical analysis of
the physiological phenomena focused here.

\section{EXPERIMENTAL EVIDENCES AND CONNECTIONS}

\subsection{$D=2$ turbulence in pure-electron plasma}

A few years ago, in 1994, Huang and Driscoll\cite{huangdriscoll} exhibited some quite interesting 
nonneutral plasma experimental results obtained in pure-electron plasma (confined in a 20 cm long and 6 
cm wide metallic cylindric Penning trap with a $10^{-10}$ torr vacuum in its interior) in the presence of 
an external  axial magnetic field (507 Gauss). In the interval 2-100 ms after every single electric shot 
(generating the electron plasma), it was observed a turbulent axisymmetric  metaequilibrium state, the 
electronic density radial distribution of which was measured. Its average (over typically 100 shots) 
monotonically decreased with the radial distance, disappearing at some radius sensibly {\it smaller} than 
the radius of the container (i.e., a {\it cut-off} was observed). The experiment was recently 
redone\cite{huangdriscoll2} under slightly modified experimental conditions (a slow external rotation 
was imposed in such a way as to compensate the small energy dissipation existing in the plasma), and 
essentially the same metaequilibrium state was observed during lapses of time as long as 27 hours, or 
even longer ! In addition to the 1994 experiment, the authors also proposed\cite{huangdriscoll} a 
phenomenological theory trying to reproduce the experimentally observed profile. Their proposal 
consisted on the optimization, for a given model, of a functional of the electron density $\rho( r)$ under 
constraints, namely conservation of total mass, angular momentum and energy. They presented four 
different attempts.  The first one ({\it Point Vortex Maximum Entropy}) consisted in optimizing, for a 
point vortex representation of the plasma, the BG entropy: it failed in reproducing the experimental data. 
The second attempt ({\it Fluid Maximum Entropy}) was essentially the same as the previous one, but 
using a fluid model for the plasma: the failure was even bigger. They assumed next that the problem 
possibly relied, not so much in the particular plasma model,  but rather in the chosen functional to be 
optimized. In their third attempt ({\it Global Minimum Enstrophy}), they turned  back to the point vortex 
model, but optimized the {\it enstrophy} instead of the BG entropy. The result was better than the two 
first attempts, but had the physically unacceptable feature of producing a {\it negative} electron density 
at sufficiently high radius. They then addressed their fourth attempt ({\it Restricted Minimum 
Enstrophy}), whose only difference with the third one was the fact of introducing an {\it out-of-the 
pocket} cut-off of the electron density at the proper value of the radius. This procedure was, finally, 
successful, and a very good first-approximation fitting was obtained ! The effort done by Huang and 
Driscoll was, on top of the high merit of a remarkable experiment, extremely pedagogical and 
elucidating: the main theoretical problem was {\it not} the model, but rather the choice of the functional 
to be optimized, i.e., the {\it statistics}.

The next important step in this story was done by Boghosian. He realized in 1995 and 
published\cite{boghosian} in 1996 that the Huang and Driscoll fourth, successful attempt {\it precisely} 
corresponds to the optimization of $S_q$ with $q=1/2$ ! Indeed, by following the recipes of the present 
generalized thermostatistics, he re-obtained, for the electron density profile, the {\it same} differential 
equation produced within the Restricted Minimum Enstrophy phenomenological theory, with the 
supplementary bonus of {\it not} having to introduce in an  ad hoc manner the necessary cut-off. Indeed, 
as already argued, all $q<1$ cases exhibit a cut-off intrinsic to the formalism, and the radial position of 
that cut-off nicely fits the experimental value.

The next step was performed  in 1997 by Anteneodo and myself\cite{anteneodoturbulence} (in fact, after 
related remarks by Boghosian himself). The Restricted Minimum Enstrophy theory is based on the 
enstrophy functional, which belongs to the general discussion of Casimir invariants; its form is in fact 
that of the order 2 Casimir invariant. Consequently, an epistemologically  conservative theoretical 
viewpoint is to appreciate Boghosian's effort as just a formal interesting remark, with no real physical 
necessity. It happens, however, that, for $r \rightarrow r_c-0$, ($r_c \equiv$ cut-off radius) the enstrophy 
theory yields $\rho( r) \propto (r_c - r)$ whereas the experimental data fit much better a {\it vanishing} 
derivative at $r_c$ ! We followed\cite{anteneodoturbulence} along Boghosian's lines and generalized his 
theory for arbitrary $q$. We obtained the generalized differential equation for $\rho(r )$ and showed that 
$\rho( r) \propto (r_c - r)^{q/(1-q)}$. Consequently, the experimental data fit better for $q$ slightly {\it 
above} $1/2$. This, together with the numerical solution of the differential equation, advanced $q \simeq 
0.55$ as a better value for satisfactory overall fitting. (Better fittings would probably demand for a model 
more sophisticated than the point vortex one used here). The conceptually important point of this 
discussion is that Casimir invariants are characterized by {\it integer} exponents (in $\rho(r )$), hence {\it 
none} of them can be related to a value of $q$ close to $0.55$. From this standpoint, the present 
formalism appears as the {\it only} satisfactory phenomenological theoretical approach available in the 
literature at the present time.

The last step of this analysis was performed very recently by Anteneodo\cite{anteneodoprivate}. Indeed, 
the calculations above recalled\cite{boghosian,anteneodoturbulence} were done by using unnormalized 
$q$-expectation values. However, as already mentioned  and used in the present review, it has been 
recently argued\cite{tsamepla} that normalized $q$-expectation values should be used instead. It is 
therefore important to check that the present discussion and results for turbulence remain essentially 
invariant. This is now done\cite{anteneodoprivate}, and it is this theory that we present in what follows.

The generalized entropy and associated constraints are given by
\begin{equation} \label{entro3}
S_q[g] \equiv 
\frac{1}{q-1} \int (g-g^q) {\rm d}^2 {\bf r} ,
\end{equation}
\begin{eqnarray} \nonumber
\int g {\rm d}^2 {\bf r}     &=& 1\;\;\;\;\;\;\;\;\;\;\;\;(mass\;conservation)
\\ \nonumber
\frac{ \int r^2 g^q {\rm d}^2 {\bf r} }{\int  g^q {\rm d}^2 {\bf r}} &=& L_q \equiv 
L\;\;\;(angular\;momentum\;conservation)
\\ \label{constrq3}
\frac{ -\frac{1}{2}
         \int \frac{\phi}{\phi^\star} g^q {\rm d}^2
         {\bf r}}{\int  g^q {\rm d}^2 {\bf r} }&=& U_q \equiv U\;\;\;(energy\;conservation),
\end{eqnarray}
where $g(r )$ is the probability distribution.
Moreover, the scaled electrostatic potential
\begin{equation}
\frac{\phi(r)}{\phi^\star} \equiv 
\frac{ \int g^q(r')G({\bf r},{\bf r}') {\rm d}^2 {\bf r^{\prime}} }{\int  g^q {\rm d}^2 {\bf r} }   
\hspace*{1cm}
\mbox{with $\nabla^2 G({\bf r},{\bf r}')=4\pi\delta({\bf r}-{\bf r}')$},
\end{equation}
 satisfies
\begin{equation}
\nabla^2 \frac{\phi}{\phi^\star} = 4\pi \frac{g^q}{\int  g^q {\rm d}^2 {\bf r}}.
\end{equation}
\\[3mm]
The constrained optimization of $S_q[g]$ 
($\delta (S_q-\alpha\int g {\rm d}^2 {\bf r}-\lambda L_q-\beta U_q)$)
now yields 
\begin{equation}
\frac{1-q\,g_q^{q-1}}{q-1}-\alpha -\frac{\lambda}{N} q r^2 \, g_q^{q-1}
+ \frac{\beta}{N} q \frac{\phi_q}{\phi^\star} g_q^{q-1} 
+ q(L\frac{\lambda}{N}-2U\frac{\beta}{N})g_q^{q-1}=0
\end{equation}
(where $ \int  g^q {\rm d}^2 {\bf r} \equiv N$)  or
\begin{equation}
\frac{g_q^{1-q}-q}{q-1}-\alpha q^{1-q} -\frac{\lambda}{N} q r^2 + 
\frac{\beta}{N} q \frac{\phi_q}{\phi^\star} 
+ q(L\frac{\lambda}{N}-2U\frac{\beta}{N})=0
\end{equation}
or, taking the Laplacian of both sides, 
\begin{equation}  \label{lapla3}
[1+\alpha (1-q)]\frac{\nabla^2 g_q^{1-q}}{q-1}-4\frac{\lambda}{N} q + 
4\pi\frac{\beta}{N^2}
 q g_q^q =0
\end{equation} 
which can be rewritten as
\begin{equation} \label{eqf3}
g''_q-q\frac{(g'_q)^2}{g_q}+\frac{g'_q}{r}=g_q^q(B^{\dag} g_q^q-A^{\dag})
\end{equation}
where $A^{\dag} \equiv 4 q \frac{\lambda}{N}/[1+\alpha (1-q)]$ and 
$B^{\dag} \equiv  4\pi q \frac{\beta}{N^2}/[1+\alpha (1-q)]$. 
\\[3mm]
Alternatively, identifying $\rho_q \equiv g_q^q/N$, we have

\begin{equation} \label{eqro3}
\rho''_q 
-\frac{2q-1}{q}\frac{(\rho'_q)^2}{\rho_q}+\frac{\rho'_q}{r}
=q  \rho_q^\frac{2q-1}{q}
 (B\rho_q-A) ,
\end{equation}
with $A \equiv A^{\dag} N^\frac{q-1}{q}$ and $B \equiv B^{\dag} N^\frac{2q-1}{q}$. This equation 
precisely is the one appearing in \cite{anteneodoturbulence}, which, for $q=1/2$, recovers that of 
\cite{boghosian}. For any chosen $q$, the values of the parameters $(A,B)$ are obtained by imposing the 
experimental values of total angular momentum and energy. This phenomenological theory has, 
therefore, only one fitting parameter ($q$). As said before, $q=1/2$ exactly reproduces the Huang and 
Driscoll's Restricted Minimum Enstrophy profile. The best overall fitting is, however, obtained for a 
value of $q$ slightly above $1/2 $.

\subsection{Solar neutrino problem}

As easily conceivable, the core of the Sun is a very complex and turbulent plasma, within which an 
enormous amount of nuclear reactions take place. Many of them constitute chains of nuclear reactions in 
which neutrinos are produced. For instance, the p-p chain is described in \cite{berezinsky}. Through a 
quite complete analysis of the production of neutrinos within the so called {\it Solar Standard Model} 
(SSM), it is possible to predict the neutrino flux onto the Earth. However, the actual flux measured in a 
variety of underground laboratories (Gallex, Sage, Kamiokande, Super-Kamiokande, Homestake) roughly 
amounts to only half of the predicted value ! This problem is currently referred to as the "solar neutrino 
problem". Two nonexclusive sources of explanation of this enigmatic discrepancy are: (i) the possible 
neutrino oscillations, which would make that only part of the predicted value would be detectable on the 
Earth; (ii) the current use of the SSM might be incorrect because it uses BG thermal statistics, which 
could be inappropriate for the solar plasma. Clayton\cite{clayton} was the first to address the second 
possibility, as far as 25 years ago ! Indeed, he assumed an hypothetic distribution of energies essentially 
given by
\begin{equation}
p(E) \propto e^{-\beta E}\;e^{-\delta(\beta E)^2}
\end{equation}
The particular value $\delta=0$ obviously recovers BG statistics. Clayton showed that a small value of 
$\delta$ ($\delta \simeq 0.01$) was enough to make the theory consistent with the experimental data that 
were available at that time. Quarati and co-workers remarked (preliminarily in 1996\cite{quarati1}, and 
in more refined calculations since then\cite{quarati2}) that, since the needed $\delta$ is very small, the 
ansatz distribution could as well be the power-law one which appears in the present formalism. By 
identifying the first corrections (to BG) of both distributions, they obtained 
\begin{equation}
\delta=\frac{1-q}{2}
\end{equation}
Consequently, values of $q$ quite close to unity are enough to fit the solar neutrino discrepancy. Once 
again, we verify the extreme efficiency that modifications of the statistics can have.

\subsection{Peculiar velocities in Sc galaxies}

From the data obtained by the Cosmic Background Explorer (COBE), it has been possible to infer the 
distribution of peculiar velocities of certain groups of spiral (Sc) galaxies (we recall that by {\it peculiar} 
velocity we mean the residual velocity after the global universe expansion velocity has been substracted). 
Bahcall and Oh\cite{galaxies} developed four theoretical attempts (namely Cold Dark Matter with 
$\Omega=0.3$ and with $\Omega=1.0$, Hot Dark Matter with $\Omega=1.0$ and Primeval Barionic 
Isotropic with $\Omega=0.3$). All the attempts were done within BG statistics. The less unsatisfactory 
fitting was obtained for the CDM model with $\Omega=0.3$. In fact, all the attempts exhibit a long tail 
towards high velocities, whereas the experimental data show a pronounced cut-off at about $500\; Km\; 
s^{-1}$. It is relevant to mention that all the models that were used had several fitting parameters, and 
nevertheless could not get rid of the tail. A fitting was then advanced\cite{galaxies2} using the present 
formalism with only two free parameters, one of them being $q$ and the other one a characteristic 
velocity. The function that was used was the $q$-generalized Maxwell distribution, essentially 
corresponding to an ideal classical gas. The quality of the fitting is quite remarkable, far better than those 
corresponding to the already mentioned four attempts. Once again, one sees that modifications of the 
statistics can be sensibly more efficient than modifications of the model. A famous example along this 
line is provided by the completely different physics associated with a gas of free fermions or of free 
bosons, i.e., a Fermi-Dirac ideal gas or a Bose-Einstein ideal gas (same model but different statistics).

\subsection{Nonlinear inverse bremsstrahlung absorption in low pressure argon plasma}

Liu et al\cite{bremsstrahlung} provided in 1994 strong evidence of the existence of non-Maxwellian 
velocity distributions in a specific plasma experiment, where low pressure argon is exposed to pulsed 
discharges. During the afterglow, measurements of the inverse bremsstrahlung of intense microwaves is 
performed. The experimental setting is such that Coulombian collisions are dominant. The experimental 
data were fitted with the following flat-topped distribution:
\begin{equation}
f(v) \propto exp[-(v/v_m)^m]
\end{equation}
with $m \ge 2$.
Souza and myself\cite{bremsstrahlung2} showed in 1997 that the same data can equally well be fitted 
with
\begin{equation}
f(v) \propto [1-(1-q)(v/v_q)^2]^{q/(1-q)}
\end{equation}
with $q \ge 1$. Furthermore, if we expand both fitting functions in the neighborhood of the Gaussian 
case, we obtain that
\begin{equation}
q=\frac{m}{2}
\end{equation}
In both fittings, the exponents $m$ and $q$ depend on the microwave power. In order to discriminate 
between the two fitting functions, quite precise and systematic experiments would be needed, in 
particular exploring the actual dependence of the results on the power.

\subsection{Cosmic background microwave radiation}

The most accurate data concerning the cosmic microwave background radiation have been obtained with 
the FIRAS (Far-infrared absolute spectrophotometer) instrument in the COBE (Cosmic background 
explorer) satellite\cite{mather}. These data are known to follow, in the $2-20\;cm^{-1}$ region, Planck's 
black-body law. In 1995, Sa Barreto, Loh and myself\cite{planck1}, as well as Plastino, Plastino and 
Vucetich\cite{planck2} (and several others since then), analyzed within what precision one is allowed to 
assume $q=1$. The result that has systematically come out from these analyses is $|q-1| < 10^{-4}$. If 
new observations were performed in the future which would be say 10 times more precise than the 
available ones\cite{mather}, this bound would be attained. Consequently, we would know better within 
what degree of confidence extensive thermostatistics can be used for this cosmological problem. If $q \ne 
1$ turns out to be clearly confirmed, it is not excluded that we would have to revise our notions about the 
structure of space-time at the appropriate scales (possibly, Planck's length). It might come out that the 
physics at that level are better described by {\it finite}-difference equations than by differential equations 
!

\subsection{Electron-positron collisions}

The electron-positron annihilation into a virtual photon and the subsequent creation of a quark-antiquark 
pair provides the cleanest environment for the hadroproduction. Each of the two initial partons begins a 
complex cascade related to the strong-coupling long-distance regime of Quantum Chromodynamics. A 
partially successful global description of the hadroproduction has been provided through a 
thermodynamical equilibrium approach, mainly that of Hagedorn in 1965\cite{hagedorn}.  This theory 
provides the following prediction:
\begin{equation}
\frac{1}{\sigma}\frac{d\sigma}{dp_T} \simeq c p_t^{3/2} \exp{-p_T/T_0}  \;\;\;\;(p_T > T_0) 
\end{equation}
where $\sigma$ is the distribution of the transverse momenta $p_T$, $T_0$ is a characteristic 
temperature which Hagedorn predicts to be independent from the electron-positron collision energy $W$ 
in the mass center referential, and $c$ is a constant. This theory fits the data quite well for small $W$, 
say $W < 10\;Gev$, but exhibits a  pronounced failure for $W$ increasing up to say $160\; Gev$. Very 
recently, Bediaga, Curado and Miranda\cite{becumi} have used, along Hagedorn's lines, the present 
generalized statistics. The results are indicated in Figs. 8 and 9. Remark that (i) $q$ varies smoothly and 
monotonically with varying $W$ (Hagedorn's theory is recovered in the $W \rightarrow 0$ limit), and (ii) 
$T_0 \simeq 0.11 \;Gev$ and practically independs from $W$ as desirable from  Hagedorn's arguments. 
These results can be considered as a strong evidence of the applicability of the nonextensive 
thermostatistics to specific anomalous systems.

\subsection{Emulsion chamber observation of cosmic rays}

Cosmic rays can be observed by using a variety of detectors (such as Pb detectors; see \cite{lattes} and 
references therein). Typically, showers of (clustered or individual) elementary particles appear which 
start at the so called vertex. These vertex are localized at various depths. The distribution of their depths 
can be measured (see\cite{wilk} and references therein for the measurements done at the Mount Pamir 
lead chambers) . This distribution was recently fitted by Wilk and Wlodarcsyc\cite{wilk2} with the 
$q=1.3$ function which emerges within the present formalism.

\subsection{Reassociation of heme-ligands in folded proteins}

In the folded conformational state, proteins might exhibit fractal effects. One such case might be the time 
evolution of the re-association of molecules that have been taken away from their natural positions. For 
instance, if $O_2$ molecules are dissociated, through light flashes, from their natural $Fe$ positions in a 
 heme protein and reach positions outside the heme pocket, they tend to start rebinding, and, for so doing, 
they might have to follow a fractal path, or be under the dynamical influence of fractal excitations (e.g., 
fractons). Anyhow, this re-association phenomenon has been lengthily studied by Frauenfelder et 
al\cite{frauenfelder}. If we define $\xi \equiv N(t)/N(0)$ where $N(t)$ is the number of molecules that 
have not yet re-associated at time $t$, the $\xi(t)$ monotonically vanishes with $t$. The results obtained 
by photo-dissociating $CO$ molecules from Sigma Type 2 sperm whale Myoglobin (Mb) dissolved in a 
glycerol-water solution are shown in Fig. 10. For times not too long, the experimental data have been 
fitted by Frauenfelder et al\cite{frauenfelder} with
\begin{equation}
\xi=(1+t/t_0)^{-n}
\end{equation}
where $t_0$ and $n$ smoothly depend on the temperature $T$. Bemski, Mendes and 
myself\cite{bemski} have argued that, within the generalized formalism, the following equation naturally 
appears:
\begin{equation}
\frac{d\xi}{dt}=-\lambda_q\;\xi^q\;\;\;\;(\lambda_q \ge 0; \;q \ge 1)
\end{equation}
Its solution is given by
\begin{equation}
\xi=\frac{1}{[1+(q-1)\lambda_q t]^{\frac{1}{q-1}}}
\end{equation}
This expression recovers, for $q=1$, the usual exponential relaxation, and reproduces the Frauenfelder 
form through the identifications $1/(q-1) \equiv n$ and $1/[(q-1)\lambda_q] \equiv t_0$. Besides 
reobtaining the Frauenfelder empiric law, the present scheme allows for a better approximation if a 
crossover is admitted. More precisely, the above differential equation can be generalized as follows:
\begin{equation}
\frac{d\xi}{dt}=-\mu_r\;\xi^r         -(\lambda_q-\mu_r)\;\xi\;\;\;\;(r \le q)
\end{equation}
The general solution involves\cite{bemski} a hypergeometric function. The fitting is shown in Figs. 10 
and 11. A detailed model which would justify the above phenomenological differential equation would 
be welcome.

\subsection{Diffusion of Hydra Vulgaris}

Upadhyaya et al\cite{arpita} are presently performing interesting experiments on {\it Hydra Vulgaris} (a 
cylindrical body column with inner and outer cells, respectively referred to as endodermal and 
ectodermal respectively) in physiological solution. The endodermal cells are more adhesive than the 
ectodermal ones. The authors have measured the velocity distribution $P(|V_y|)$ of the ``vertical" 
component of the velocity during the diffusion of endodermal Hydra cells in an ectodermal aggregate. 
The results are presented in Fig. 12, where the velocity unit is $10^{-6}m/hour$ and the probability is 
represented by the histogram of the number of counts. These results were fitted with
\begin{equation}
P(|V_y|)=\frac{a}{(1+b |V_y|^2)^c}
\end{equation}
with the values of $(a,b,c)$ indicated in the figure. Through the identification
\begin{equation}
a=P(0);\;\;
b=(q-1)/V_0^2;\;\;
c=\frac{q}{q-1}
\end{equation}
we precisely have the law which emerges within the present formalism, namely
\begin{equation}
P(|V_y|)=\frac{P(0)}{[1+(q-1)( V_y/V_0)^2]^{q/(q-1)} }
\end{equation}
with $q=1.53$. The next desirable step of course is to formulate a specific model for Hydra which would 
lead to this law, but this remains to be done. 

\subsection{Citation of scientific papers}

An interesting study was recently done by Redner\cite{redner}, in which the statistics of citations of 
scientific papers is focused. He exhibited the number $N(x)$ of papers which have been cited $x$ times 
for two long series, namely one (6 716 198 citations of 783 339 papers) from the Institute of Scientific 
Information (ISI) and another one (351 872 citations of 24 296 papers) from the Physical Review D 
(PRD). As expected, in both examples, $N(x)$ monotonically decreases with $x$. Redner fitted the 
(relatively) low-$x$ data with a stretched exponential of the form 
\begin{equation}
N(x)=N(0)\;e^{-(x/x_0)^{\beta}}
\end{equation}
with $\beta = 0.44$ and $0.39$ for the series ISI and PRD respectively. Also, he remarked that the large-
$x$ data exhibit a power law, namely close to $\propto 1/x^3$. He argues that this different functional 
behavior for low and large values of $x$ must reflect different phenomenologies in these two regimes. In 
contrast with this viewpoint, Albuquerque and myself\cite{marcio} argue that this is not necessarily so 
since the data can be quite satisfactorily fitted with a {\it single} function, namely
\begin{equation}
N(x)=\frac{N(0)}{[1+(q-1) \lambda x]^{q/(q-1)}}
\end{equation} 
with $q=1.53$ and $1.64$ for the series ISI and PRD respectively: see Fig.13 The satisfactory quality of 
the fittings is, after all, not so surprising, since we have mentioned earlier in this paper the 
connection\cite{denisov} of this formalism with the Zipf law.
   
\subsection{Electroencephalographic signals of epilepsy}

It is since long known that the analysis of signals can be done within formalisms which use entropic 
forms. One such application has been recently done on EEG records of epilleptic humans and 
turtles\cite{epilepsia}. The simultaneous use of wavelet-based multiresolution analysis including the 
nonextensive entropy $S_q$ leads to signals whose interpretation can be clinically neat and 
pharmacologically convenient. The authors of this novel processing suggest perspectives for building up 
automatic detection devices.

\subsection{Cognitive psychology}

The development of  artificial neural networks and their connections with statistical mechanics (e.g., the 
Hopfield model for associative memory) makes quite natural the approach of cognitive problems with the 
present nonextensive formalism. Within this philosophy, we performed\cite{alexandra} an experiment of 
learning/memorization (of $5 \times 5$ and $7 \times 7$ square matrix having circles and crosses 
randomly distributed once for ever) with students of the University level; 150 students were interviewed, 
the first 30 in order to optimize the experimental protocole, and the other 120 to make the measurements 
of the time-evolution of the total amounts of errors when the original matrix was successively shown and 
hidden. The average results were then fitted with those obtained, for the same task, with a learning 
machine\cite{cordobamachine} having a perceptron architecture and an internal dynamics based on the 
Langevin equation\cite{langevin} generalized by Stariolo to arbitrary $q$.  The (average) learning time 
of the machine turned out to monotonically increase with $q$, exhibiting a practically divergent 
derivative at $q=1$. The best human-machine fit occurred for $q$ slightly above unity. More 
experiments and comparisons along these lines would be very welcome. Indeed, they would help better 
understanding some cognitive phenomena, on one hand, and could generate efficient machines for 
specific tasks, on the other.

\subsection{Turbulence and time evolution of financial data}

In 1996 Ghashghaie et al\cite{turbofinance} compared financial data with those obtained from turbulent 
behavior and showed very similar behaviors when appropriate scalings are used. Ramos et al\cite{ramos} 
have recently shown that all these data can be satisfactorily fitted with the functional form which emerges 
from the present formalism. Olsen and Associates data containing bid-ask quotes for US dollar-German 
mark exchange rates (1,472,241 records) are presented in  Fig. 14 (probability density $P_{\Delta 
t}(\Delta \pi)$ of price changes; $\Delta \pi = \pi(t)-\pi(t+\Delta t)$ with $\Delta t= 
640s,\;5120s,\;40960s,\;163840s$ from top to bottom in the figure). The turbulent flow 
data\cite{chabaud} are presented in Fig. 15 (probability density $P_{\Delta r}(\Delta v)   $ of velocity 
differences; $\Delta v =v( r)-v(r+ \Delta r)$ for spatial scale delays $\Delta r = 3.3 \eta,\;18.5\eta,\; 138 
\eta,\; 325 \eta$ from top to bottom in the figure, where $\eta$ is the Kolmogorov scale, i.e., the critical 
limit for occurence of viscous dissipation). All these data exhibit a slight left-right assymmetry, which 
has been taken into account by Ramos et al: they used the same $q$ for both sides but different widths. 
Needless to say that specific models leading to these fitting functions are very welcome.

\section{COMPUTATIONAL EVIDENCES AND CONNECTIONS}

\subsection{Thermalization of a hot gas penetrating in a cold gas}

In 1991, Waldeer and Urbassek\cite{waldeer} made, assuming $d=3$ Newtonian mechanics, a 
computational simulation in which a certain amount of high energy particles penetrate into a cold gas and 
are thermalized through the interactions between molecules. The cold gas is initially put at BG thermal 
equilibrium at temperature $T_C$. The high energy particles at time $t=0$ are randomly distributed in 
energy at a quite high energy per particle. The interaction potential was assumed to be hard sphere at 
short distances and decreasing, at long distances, like $r^{-\alpha}$. They analyzed three typical 
situations, one with $\alpha \rightarrow \infty$, hence well above $d$ (i.e., very short range interactions), 
the second one with $\alpha=4$ (i.e., short range interactions), and the last one with $\alpha=8/3$, which 
is below $d$ (i.e., long range interactions). In their simulation, they follow the time evolution of the 
energy distribution of the hot particles. After a transient, this distribution evolves with a regular pattern. 
For $\alpha > d$, this pattern basically is  the BG distribution with a temperature $T(t)$ which gradually 
approaches $T_C$ from above (with $lim_{t \rightarrow \infty} T(t) = T_C$), in other words, through 
curves which approximatively are straight lines in a log-linear plot. For $\alpha<d$, this approximation 
occurs through curves which are close to straight lines... in a log-log plot ! (Notice that the curvature in 
log-log plots tends to be {\it upwards} for $\alpha<d$, whereas it is {\it downwards} for $\alpha >d$; see 
Figs. 1, 2 and 3 of \cite{waldeer}).  This power-law behavior is typical of $q>1$. This pecualiarity was 
invoked by Koponen\cite{koponen} in 1997 as a justification for using the present generalized formalism 
to discuss electron-phonon relaxation in ion-bombarded solids if the interactions are long-ranged. A study 
like that of Waldeer and Urbassek\cite{waldeer} which would systematically address the details of that 
thermalization by gradually varying $\alpha$ across $d$ is missing and would certainly be very welcome.

\subsection{Long-range classical Hamiltonian systems: Static properties}

Let us focus here on what we refer to as {\it weak} violation of BG statistics. We use this expression to 
distinguish it from what we call {\it strong} violation of BG statistics. Both of them lead to nonextensive 
quantities, but, whereas the strong violation concerns $q \ne 1$, the weak one concerns $q=1$ 
calculations. To make all this explicit we shall here focus on classical systems, i.e., all observables are 
assumed to commute. Let us consider the following paradigmatic Hamiltonian:
\begin{equation}
{\it H} = \frac{1}{2m}\sum_{i=1}^N p_i^2 + \sum_{i \ne j} V(r_{ij})
\end{equation}
where $m$ is a microscopic mass, $\{p_i,r_i\}$ are the $d$-dimensional linear momenta and positions 
associated with $N$ particles, and $r_{ij} \equiv r_j-r_i$. A typical situation is that of a finite confined 
system but, if some care is taken, the system could as well be thought  of as having periodic boundary 
conditions. To be specific, let us  assume
\begin{equation}
V(r_{ij}) =  \frac{A}{r_{ij}^{12}} -\frac{B}{r_{ij}^{\alpha}}\;\;\;(A>0, B>0, 0 \le \alpha <12)
\end{equation}
where, in order to avoid any singularity at the origin (for any dimension $d$ not exceedingly high), we 
have assumed, for the repulsive term, the Lennard-Jones exponent 12. What we desire to focus on in the 
present discussion is possible singularities associated with infinite distances, i.e., the effects of {\it long}-
range (attractive) interactions. The case $(\alpha,d)=(6,3)$ precisely recovers the standard Lennard-Jones 
fluid; the case $(\alpha,d)=(1,3)$ is asymptotically equivalent to Newtonian gravitation; the case 
$(\alpha,d)=(d-2,d)$ is asymptotically equivalent  to $d$-dimensional gravitation (i.e., the one associated 
with the solutions of the $d$-dimensional Poisson equation); the case $(\alpha,d)=(3,3)$ basically 
reproduces the distance dependance of permanent dipole-dipole interaction. The range of the (attractive) 
interaction increases when $\alpha$ decreases; $\alpha \rightarrow 12$ corresponds to very short-ranged 
interactions, whereas $\alpha=0$ corresponds to the situation of the Mean Field Approximation, where 
every particle (attractively) interacts with every other with the {\it same} strength, in all occasions.

A typical quantity to be calculated within BG statistics is the following one (basically related to the 
$T=0$ internal energy per particle):
\begin{equation}
\int_1^{\infty}dr\;r^{d-1}\;r^{-\alpha}
\end{equation}
where the distances $r$ have been expressed in units of a characteristic length of the problem. We 
immediately verify that this integral {\it converges} if  $\alpha>d$, and {\it diverges} if $0 \le \alpha \le 
d$. Consequently, thermodynamic calculations in the $0 \le \alpha \le d$ case have to be done with some 
care, and not blindly following the standard rules associated with BG statistics (i.e., $q=1$). It is in this 
sense that we use the expression ``weak" violation of BG statistics. The care to which we refer is the fact 
that we have to strictly consider the {\it finite} size of the physical system. Consistently, a relevant 
quantity that emerges naturally is
\begin{equation}
N^* \equiv d\int _1^{N^{1/d}}dr\;r^{d-1}\;r^{-\alpha} = \frac{N^{1-\alpha/d}-1}{1-\alpha/d}
\end{equation}
We can check that, in the $N \rightarrow \infty$ limit, we have
\begin{equation}
N^* \sim \left \{
\begin{array}{c}
\!\!\!\!\!\!\! \frac{1}{\alpha/d -1} \quad \mbox{if}\quad \alpha/d >1; \\
\!\!\!\!\!\!\! \ln\; N \quad
\mbox{if} \quad \alpha/d =1; \\
\frac{N^{1-\alpha/d}}{1-\alpha/d}\quad \mbox{if} \quad
0 \le \alpha/d < 1 .
\end{array}
\right.
\end{equation}
As it will become transparent later on, what these regimes imply is that the system is {\it extensive} for 
$\alpha/d > 1$ (hence standard thermodynamics apply), whereas it is {\it nonextensive} for $0 \le 
\alpha/d \le 1$, and special scalings become necessary \cite{tsallisfractals,jund} in order to have both a 
mathematically well posed problem, and a physical unfolding (or qualification) of the nonextensive 
region. The 
$\alpha/d > 1$ regime has since long been analyzed\cite{fisher}, and it is well known that extensivity (or, 
{\it stability}, as also referred to) is lost for $\alpha/d \le 1$. However, to the best of our knowledge, the 
scalings associated with the quantity $N^* $,  as well as its numerically efficient collapsing properties, 
were introduced for the first time by Jund et al\cite{jund} in 1995.

A quantity related to $N^*$, namely $\tilde{N}$  turns out to be even more convenient. It is defined 
through\cite{openqu}
\begin{equation}
\tilde{N} \equiv N^* +1 = \frac{[N^{1-\alpha/d}-\alpha/d]}{1-\alpha/d}
\end{equation}
In the $N \rightarrow \infty$ limit, we have
\begin{equation}
\tilde{N} \sim \left \{
\begin{array}{c}
\!\!\!\!\!\!\! \frac{\alpha/d }{\alpha/d -1} \quad \mbox{if}\quad \alpha/d >1; \\
\!\!\!\!\!\!\!  \ln\; N \quad
\mbox{if} \quad \alpha/d =1; \\
\frac{N^{1-\alpha/d}}{1-\alpha/d}\quad \mbox{if} \quad
0 \le \alpha/d < 1 .
\end{array}
\right.
\end{equation}
In the limit $\alpha/d \rightarrow \infty$, $\tilde{N} \rightarrow 1$; in the limit $\alpha/d \rightarrow 
1+0$, $\tilde{N} \sim 1/(\alpha/d-1)$; in the limit $\alpha/d \rightarrow 0$, $\tilde{N} \sim N$. Roughly 
speaking, $\tilde{N}$ characterizes the {\it effective} number of neighbors that can be associated with a 
given particle. This is the convenience to which we referred above.

We are ready now to present the kind of size-scalings we expect to be necessary for thermodynamically 
describing a generic classical Hamiltonian system with the type of interactions above mentioned. Let us 
focus on a simple fluid, and start with the standard case, i.e., $\alpha>d$. Its Gibbs energy $G(T,p,N)$ is 
given by
\begin{equation}
\frac{G(T,p,N)}{N} \sim \frac{U(T,p,N)}{N} -T\;\frac{S(T,p,N)}{N}+p\;\frac{V(T,p,N)}{N}
\end{equation}
where $U,\;S,\;V,\;N,\;T$ and $p$ respectively are the total internal energy, total entropy, total volume, 
total number of particles, temperature and pressure. In the $N \rightarrow \infty$ limit, we obtain
\begin{equation}
g(T,p)=u(T,p)-T\;s(T,p)+p\;v(T,p)
\end{equation}
where the corresponding densitary variables have been introduced.

In contrast with the above, if we have $0 \le \alpha \le d$, the scalings are different, namely
\begin{equation}
\frac{G(T,p,N)}{N\;\tilde{N}} \sim \frac{U(T,p,N)}{N\;\tilde{N}} -
\frac{T}{\tilde{N}}\;\frac{S(T,p,N)}{N}+\frac{p}{\tilde{N}}\;\frac{V(T,p,N)}{N}
\end{equation}
Consistently we have
\begin{equation}
g(\tilde{T},\tilde{p})=u(\tilde{T},\tilde{p})-
\tilde{T}\;s(\tilde{T},\tilde{p})+\tilde{p}\;v(\tilde{T},\tilde{p})
\end{equation}
where
\begin{equation}
\tilde{T} \equiv \frac{T}{\tilde{N}};\;\;\;\;\tilde{p} \equiv \frac{p}{\tilde{N}}
\end{equation}
These equations recover the previous ones, i.e., those associated with the $\alpha>d$, as a particular case. 
Indeed, for $\alpha>d$, $\tilde{N}$ becomes a constant. 

So, we see that long range interactions have important thermodynamical consequences, namely

(i)  the {\it energy} quantities ($G,\;U$, which normally appear alone) that were {\it extensive} for 
$\alpha>d$ {\it loose} their extensivity;

(ii)  the {\it non-energy} quantities ($S,\;V$, which normally appear in canonical pairs with intensive 
quantities) that were {\it extensive} for $\alpha>d$ {\it preserve} their extensivity;

(iii) the control parameters ($T,\;p$) that were {\it intensive} for $\alpha>d$ {\it loose} their intensivity. 

Consistently, to have mathematically well defined and physically useful equations of states and related 
quantities, everything must refer to {\it finite} quantities, hence, we must express all relations with the 
above rescaled variables. This does {\it not} imply that thermal equilibrium occurs through sharing equal 
values of $\tilde{T}, \;\tilde{p}$, etc. The zero-th principle of thermodynamics appears to hold in the 
usual way, {\it even if we have long range interactions in the system.} Although we have illustrated these 
features on a fluid, it is clear that the same considerations hold for all types of thermodynamical systems 
(magnets, dielectric substances, elastic solids, etc).

Two particular remarks must be made at this point:

(i)  When every element of the system equally interacts with each other (i.e., $\alpha=0$), 
$\tilde{N}=N$, and consequently $\tilde{T}=T/N$. In what concerns the thermostatistical approach of a 
system, this is equivalent to dividing the microscopic coupling constants by $N$, a familiar feature that is 
artificially imposed in {\it all} Mean Field calculations. We have used the word ``artificial" because, 
whenever $\alpha \le d$, the Hamiltonian which includes the {\it microsopic} interactions {\it indeed is 
nonextensive!} (and so is $U$). To divide the two-body coupling constants by $N$ when $\alpha=0$ (or, 
by $\tilde{N}$, when $\alpha>0$) certainly is an artificial manner of forcing to be extensive a 
Hamiltonian which physically is not. This practice is traditionally frequent among magneticiens (who 
divide $J$ by $N$), but certainly not among astronomers, who normally do not even think about what 
would be a very strange way of renormalizing the gravitational constant $G$ !

(ii) If a singularity (for example, a critical phenomenon) occurs under particular physical conditions, it 
must generically occur at {\it finite} values of  $\tilde{T},\;\tilde{p}$, etc, and {\it not} at finite values of 
$T,\;p$, etc. Let us illustrate this for the simple case of a critical temperature: $\tilde{T}_c$ must be 
finite, hence, if $\alpha/d>1$, it must be $T_c(\alpha,d) \propto 1/(\alpha/d-1)$ This implies that $T_c$ 
must {\it generically} diverge for {\it all} classical systems at the extensive-nonextensive frontier. More 
precisely
\begin{equation}
T_c(\alpha,d) \sim \frac{A(d)}{\alpha/d-1} \;\;\;\;(\alpha/d \rightarrow 1+0)
\end{equation} 
where $A(d)$ is a system-dependent finite constant. In fact, let us anticipate that this precise behavior has 
been observed in the systems available in the literature\cite{jund} ($d=2$ and $d=3$ Lennard-Jones-like 
fluids, $d=1$ and $d=2$ Ising and Potts ferro- and antiferromagnets, etc), with no exception. To illustrate 
the connection between fluid models like the extended Lennard-Jones one above considered, and 
localized spin systems, let us briefly focus on the Ising ferromagnet. The simplest long-range $N$-spin 
cubic-lattice Hamiltonian of this kind is given by\cite{hiley}
\begin{equation}
{\it H}=-J\sum_{i \ne j}\frac{S_iS_j}{r_{ij}^{\alpha}}\;\;\;\;\;(J>0;\;\alpha \ge 0;\;S_i = \pm 1\; \forall i )
\end{equation}
where, for $d=1$, $r_{ij}=$1, 2, 3, ...; for $d=2$, $r_{ij}=1,\; 2^{1/2},\; 2, ...$; for $d=3$, $r_{ij}=1, 
\;2^{1/2},\; 2^{1/3},\; 2, ...\;$; and so on. Clearly, the limit $\alpha \rightarrow \infty$ recovers the first-
neighbor $d$-dimensional spin 1/2 ferromagnet, whereas $\alpha=0$ corresponds to the Mean Field 
Approximation. For this model, $kTc(\alpha,d)/J$ diverges for $0 \le \alpha/d <1$, decreases for  
$\alpha/d$ increasing above unity, and approaches the first-neighbor value (0 for $d=1$, 2.269... for 
$d=2$, etc) for $\alpha/d \rightarrow \infty$. Also, $kT_c/J \sim A(d)/[(\alpha/d)-1]$ in the $\alpha/d 
\rightarrow 1+0$ limit. The introduction of $\tilde{T}$ nicely enables the {\it unfolding} of the region 
where $T_c$ diverges. Indeed,  $k\tilde{T}_c/J \equiv kT_c/(J \tilde{N})$ is {\it finite} in {\it both} 
extensive ($\alpha/d >1$) and nonextensive ($0 \le \alpha/d \le 1$) regions, thus providing an  
enlightening unification. Finally, let us mention that, it seems that all equations of states (e.g., $lim_{N 
\rightarrow \infty}M(T,N)/N$ can be, for all $\alpha/d \le 1$, mapped into that associated with the 
Molecular Field Approximation. This simplifying feature appears to hold {\it only} for the static 
thermodynamic properties, and not for the dynamical ones, as will become clear later on (in Subsection 
V.G).

To close this subsection, let us emphasize that what we have been focusing on here is what we refer to as 
the {\it weak} violation of BG statistical mechanics (see Fig. 4). These are  analytical or Monte Carlo 
$q=1$ calculations (i.e., the energy distribution obeys the Boltzmann factor), {\it but} the variables must 
be scaled with $\tilde{N}$, which is not at all necessary for the standard, short-ranged interacting 
systems.

\subsection{Long-range tight-binding systems}

In the previous subsection we addressed classical systems. It is clear, however, that similar 
nonextensivity is expected to emerge in quantum systems if long-range interactions are present. One such 
Hamiltonian is the tight-binding-like which follows\cite{hugo,lisajose}:
\begin{equation}
{\it H} = \sum_{i=1}^N \epsilon_i c_i^+c_i + \sum_{i,j \ne i}\frac{V}{r_{ij}^{\alpha}}c_i^+ c_j
\end{equation}
where $c_i^+$ and $c_i$ are the creation-annihilation operators associated with electrons on site $i$, the 
$\{\epsilon_i\}$ are the on-site energies, and $V$ is the inter-site energy. The $T=0$ electron diffusive 
properties corresponding to this Hamiltonian exhibit a variety of anomalies intimately related to 
$\tilde{N}$, as preliminary shown by Nazareno and Brito\cite{hugo} and studied with more details in 
\cite{lisajose}.

\subsection{Granular systems}

In 1995, Taguchi and Takayasu\cite{granular} simulated a vertically vibrated bed of powder with 
inelastic collisions and studied the distribution of horizontal velocities. In the lower layers (so called {\it 
solid phase}) they observed a standard Maxwellian (Gaussian) distribution of velocities. The situation 
was sensibly different in the upper layers (so called {\it fluidized phase}). Indeed, there the distribution 
was a Student's t-distribution, precisely the one appearing in Eq. (42) with $q=3$, and assuming an 
energy proportional to the $(velocity)^2$ (together with $d=r=2$, hence a constant density of states). 
This anomaly must be related to the fractal-like granular clusterization which occurs in real 
space\cite{granular2} but a deep analysis would be welcome. Also would further simulations, for 
instance of the cooling type. Studies of such computational models, either externally forced or just left to 
their own isolated evolution, can provide important physical insights, especially if quantities like the 
energy distribution, the Lyapunov spectrum (or at least its maximum value) or possible multifractality are 
focused on.

\subsection{$d=1$ dissipative systems}

One-dimensional maps constitute the simplest systems which might present chaos. Basically they consist 
of the following recurrent equation:
\begin{equation}
x_{t+1} = h(x_t;a)\;\;(t=0, 1, 2, ...)
\end{equation}
where $h(x;a)$ is a rather simple nonlinear function, and $a$ is a control parameter. Typically, both $x$ 
and $a$ are real numbers (but higher-dimensional situations are of course possible, and frequently 
studied).
The logistic map, for instance, exhibits this structure. Typically, for $a<a_c$, the system exhibits simple 
orbits, the attractor being a cycle whose number of elements is finite. For $a>a_c$, the system can 
exhibit attractors with an infinite number of elements. The value $a_c$ is the critical one, usually 
referred to as the {\it chaos threshold}; the associated attractor typically constitutes a {\it multifractal} 
characterized by the so called {\it multifractal function} $f(\alpha_H)$, where $\alpha_H$ is the Holder 
exponent. The $f(\alpha_H)$ function is generically concave, attains its maximum at a value of 
$\alpha_H$ in the interval $[\alpha_H^{min},\alpha_H^{max}]$ and this maximum equals the {\it 
fractal} or {\it Haussdorf} dimension $d_f$. An important feature of this type of maps is the sensitivity to 
the initial conditions (and, of course, the rounding at any intermediate calculation). More precisely, if we 
note $\Delta x(0)$ a small variation in the initial condition $x_0$, and follow its time evolution  $\Delta 
x(t)$, we can define the {\it sensitivity function} $\xi(t)$ as 
$\xi (t) = lim_{\Delta x(0) \rightarrow 0} \frac{\Delta x(t)}{\Delta x(0)}$.
At most values of $a$, $\xi(t)$ satisfies $d\xi / dt = \lambda_1 \xi$, hence 
\begin{equation}
\xi=\exp{\lambda_1\;t}
\end{equation}
where $\lambda_1$ is the so called {\it Lyapunov exponent}. If $\lambda_1>0$ ($\lambda_1<0$) the 
system is said {\it strongly sensitive (insensitive)} to the initial conditions. The $\lambda_1=0$ 
possibility can also occur and is referred to as the {\it marginal} case. In this situation $\xi(t)$ 
satisfies\cite{pesin,logistic,marcelo} $d \xi  / dt = \lambda_q\;\xi^q$, hence
\begin{equation}
\xi= [1+(1-q) \lambda_q\;t]^{\frac{1}{1-q}}
\end{equation}
which recovers Eq. (25) as the $q=1$ case. Two $\lambda_1=0$ possibilities exist, namely $q<1$ with 
$\lambda_q>0$ ({\it weakly sensitive} to the initial conditions), and $q>1$ with $\lambda_q<0$ ({\it 
weakly insensitive} to the initial conditions). For instance, the logistic map exhibits $q=1$ for almost all 
values of $a$ but exhibits $q<1$ at the chaos threshold and $q>1$ at every doubling-period as well as 
tangent bifurcations. Moreover, it has been shown that a large class of such systems (for which 
$f(\alpha_H^{min})=f(\alpha_H^{max})=0$) verify, at the chaos threshold, the following scaling 
law\cite{marcelo}:
\begin{equation}
\frac{1}{1-q} = \frac{1}{\alpha_H^{min}}- \frac{1}{\alpha_H^{max}}
\end{equation}
This is a fascinating relation. Indeed, its left-hand member concerns the dynamics of the sensitivity to 
initial conditions of the map, whereas its right-hand member concerns pure, though nontrivial, geometry. 
Under what precise mathematical conditions does it hold? How should it be generalized in order to also 
cover the standard case of Euclidean geometry ($\alpha_H^{min}=\alpha_H^{max}=d_f=1$) for which 
one expects $q=1$? (Should we also consider simultaneously 
$f(\alpha_H^{min})=f(\alpha_H^{max})=1$ ?; Could Eq. (115) be generalized into say  $ 1/(1-q)=1/[ 
\alpha_H^{min}-f(\alpha_H^{min})] - 1/[ \alpha_H^{max}-f(\alpha_H^{max})]  $, the Euclidean case 
thus corresponding to a special limit of the type $ \alpha_H^{min}/f(\alpha_H^{min})= 
\alpha_H^{max}/f(\alpha_H^{max})=q=1$ ?). What happens for two- or more-dimensional maps? What 
happens if, instead of maps, we have ordinary (or even partial) differential equations? To answer all these 
questions, computational effort is invaluable. 

Before closing this Subsection it is mandatory to clarify what the index $q$ appearing in the differential 
equation yielding Eq. (114) has to do with the one appearing in the present generalized entropy. In fact, 
they are one and the same, and the connection is established through the so called Pesin equality or 
identity. Let us illustrate the basic ideas on the logistic map herein considered. Assume that we make a 
partition of the $x$ interval into a large number $M$ of equally small windows, chose arbitrarily one of 
those windows and randomly put a large number $N$ of points inside. We then calculate (by using the set 
of  $M$ probabilities corresponding to the ratios of numbers of points belonging to each window) the 
$t=0$ value of the BG entropy $S_1(0)$, which is going to be very close to zero (strictly zero in the 
$(M,N) \rightarrow (\infty,\infty)$ limit). We then allow each of the $N$ points to evolve according to 
the logistic map until an attractor is achieved. The entropy $S_1(t)$ will grow with $t$ until arrival to a 
saturation value $S_1(\infty)$ which depends on $(M,N)$ (necessarily $lim_{N \rightarrow \infty} 
S_1(\infty) < \ln\;M$). In the $M \rightarrow \infty$ limit, the growth of $S_1(t)$ is in fact linear (see, 
for instance, \cite{baranger}, which enables the following characterization of the so called Kolmogorov-
Sinai entropy:
\begin{equation}
K_1=lim_{t \rightarrow \infty} lim_{M \rightarrow \infty}  lim_{N \rightarrow \infty}  \frac{S_1(t)}{t}
\end{equation} 
Quite generically, the Pesin {\it inequality} holds, which states (for one-dimensional nonlinear dynamical 
systems) that
\begin{equation}
K_1 \le \lambda_1 \;\;if\;\lambda_1>0
\end{equation}
and $K_1=0$ otherwise. Since we are only presenting a sketch of the situation, let us from now on 
address those particular systems for which the {\it equality} holds\cite{hilborn}. For those it is
\begin{equation}
K_1 = \lambda_1 \;\;if\;\lambda_1>0
\end{equation}
and $K_1=0$ otherwise. This type of analysis is convenient either if we have simple orbits (i.e., strong 
insensitivity to the initial conditions, i.e., for $\lambda_1 < 0$) or if we have strong chaos (i.e., strong 
sensitivity to the initial conditions, i.e., $\lambda_1>0$). But this analysis is a very poor one at say the 
edge of chaos, where $\lambda_1=0$ and we have weak sensitivity to the initial conditions. It is to unfold 
this type of situation that $S_q$ becomes extremely useful. Let us show how. At the chaos threshold we 
have $K_1=\lambda_1=0$. But if we follow the same procedure we just described for calculating $K_1$, 
but using instead $S_q(t)$, an interesting phenomenon can be revealed, which we describe now. Let us 
first define the following generalized Kolmogorov-Sinai entropy:
\begin{equation}
K_q=lim_{t \rightarrow \infty} lim_{M \rightarrow \infty}  lim_{N \rightarrow \infty}  \frac{S_q(t)}{t}
\end{equation}
A value $q^*$ is generically expected to exist\cite{guerberoff} such that, for $q>q^*$ ($q<q^*$), $ 
K_q=0$ ($K_q$ diverges),  and $K_{q^*}$ is {\it finite} ! Furthermore, it can be argued\cite{pesin} that 
the above Pesin equality can be generalized as follows:
\begin{equation}
K_q = \lambda_q \;\;if\;\lambda_q>0
\end{equation}
and $K_q=0$ otherwise. It is through this important type of (in)equality that the connection emerges 
between $S_q$ and the power-law time-dependence of the sensitivity to the initial conditions. The 
particular value $q^*$ above described is what was numerically calculated in 
\cite{pesin,logistic,marcelo}, and satisfies the scaling (115)

Some of the above statements can be trivially checked with the logistic map at its chaotic region (i.e., for 
$\lambda_1>0$). We know in that case that $S_1(t) \propto t$, hence (assuming the simple case of 
equiprobability) the total number of possibilities $W(t)$ grows {\it exponentially} with $t$. For any 
$q>1$, $S_q(t)$ is {\it always} bounded, then $K_q$ necessarily vanishes. For any $q<1$, $S_q(t)$ 
grows like the $1/(1-q)$ power of $W(t)$, which in turn, as said before, grows exponentially with $t$, 
hence necessarily $K_q \rightarrow \infty$. We conclude that $q^*=1$ ! The same picture is expected to 
hold for weak chaos.

\subsection{Self-organized criticality}

In the previous example, fine tuning (e.g., $a=a_c$) is necessary to observe the anomalous ($q \ne 1$) 
behavior. Let us address dissipative systems with many degrees of freedom, very particularly those which 
do not need fine tuning. Would robust systems like those exhibiting self-organized 
criticality\cite{perbak} (SOC) also present $q \ne 1$ behavior? The answer is {\it yes}, as it has been 
clearly exhibited in at least three computational systems, namely the Bak-Sneppen model for biological 
evolution, the Suzuki-Kaneko model for imitation games and the Bak-Tang-Wiesenfeld model for 
sandpiles \cite{tamarit}. In these systems, the Hamming distance plays the role played by $\xi$ in the 
previous ones. Also, the relevance of the order  of the $t \rightarrow \infty$ and  $N \rightarrow \infty$ 
limits has been exhibited. Like in the conjectural Fig. 4, the $q \ne 1$ behavior is observed only in the 
$ lim_{t \rightarrow \infty} lim_{N \rightarrow \infty}$ order. On what model ingredients does $q$ 
depend? Is a taxonomy in universality classes analogous to that of standard critical phenomena possible? 
Is a multifractal $f(\alpha_L)$ function hidden somewhere? Does a scaling law like that of Eq. (115) still 
hold? Again, additional computational effort is very welcome.

\subsection{Long-range classical Hamiltonian systems: Dynamic properties}

Let us finally focus on the ``heart" of statistical mechanics, the dynamics of the systems on which 
Boltzmann himself was meditating, namely the Hamiltonian systems with many degrees of freedom. 
Although lots of interesting quantum nonextensive phenomena must exist, here we shall restrict ourselves 
to the classical ones. We expect them to be able to provide nonextensive anomalies in a kind of pure, or 
simpler manner. Since a classical canonical Hamiltonian must satisfy the Liouville theorem, the 
Lyapunov spectrum must be symmetric with respect to zero, the corresponding eigenvalues being 
necessarily coupled in pairs of positive and negative values with the same absolute value. Consequently, 
to study the sensitivity to the initial conditions it suffices to study the maximum Lyapunov exponent. If it 
is positive, the system will generically be {\it strongly} chaotic, and will therefore easily satisfy the 
ergodic/mixing hypothesis (equality of time and ensemble averages). If, however, the maximum 
Lyapunov exponent vanishes, the entire spectrum will necessarily vanish, hence the system will be, at 
most, {\it weakly} chaotic, and will therefore have difficulties in satisfying the ergodic/mixing 
hypothesis, at least at not extremely large times (reflecting the macroscopic size of the system). The 
$d=1$ coupled planar rotators $N$-body model with a two-body coupling constant proportional to 
$1/r^{\alpha}$ ($r \equiv $ distance between two given rotators) has been recently studied (for 
$\alpha=0$ in \cite{latora}, and, for $\alpha \ge 0$ in \cite{anteneodo}) in the microcanonical ensemble. 
It has been established that, above a critical (conveniently normalized) total energy, the maximum 
(conveniently normalized) Lyapunov exponent is, in the $N \rightarrow \infty$ limit, positive (zero) for 
$\alpha>1$ ($0 \le \alpha \le 1$). More precisely, this maximum Lyapunov exponent is proportional to 
$1/N^{\kappa}$ where $\kappa(\alpha)$ appears to be a monotonic function which decreases from 
$\kappa(0)$ to zero while $\alpha$ increases from $0$ to $1$, and remains zero for all $\alpha \ge 1$). It 
must be recalled that it is only for $\alpha >1$ that the standard BG prescriptions provide {\it finite} 
integrals in the relevant calculations. If we were to discuss the $d$-dimensional version of the same 
model, we would certainly have $\kappa(\alpha,d)$. It is certainly possible that it is $\kappa(\alpha/d)$, 
and it would be interesting to check such a hypothesis.  If we were to consider not planar (like the XY 
ferromagnet) but rather three-dimensional (like the Heisenberg ferromagnet) rotators, would $\kappa$ be 
insensitive to that, or it would depend on the specific model? What would happen if, instead of localized 
rotators, we were to consider a long-ranged Lenard-Jones-like fluid, or $d$-dimensional gravitation? All 
these questions are certainly interesting, and worthy of further computational efforts.

In a recent paper, an essentially $\alpha=0$ model was considered\cite{antoni}, and, under certain 
circumstances, a crossover was found between anomalous (at times smaller than $\tau(N)$) and normal 
(at times larger than $\tau(N)$) diffusion, with $\tau(N) \propto N$. What happens if $\alpha>0$? Does 
$\tau$ scale like $\tilde{N} \equiv N^*(1+\alpha/d) =   (N^{1-\alpha/d}-1)(1+\alpha/d) / (1-\alpha/d)$ ? 
What happens for other models? The behavior observed for the particular model that was studied is 
consistent with Fig. 4. But is it exactly that conjecture that is going on? Only the study of the energy 
distributions (of single particles or of relatively large subsystems of the $N$-body system) themselves 
can provide the answer. What about the distributions of velocities? Are they Maxwellian (i.e., Gaussian) 
for $\alpha/d>1$ and non-Maxwellian otherwise? Are they Levy's or Student's t-distributions for 
$\alpha/d \le 1$? If so, what is the dependence of $q(\alpha,d)$ ? Maybe $q(\alpha/d)$ ? Are the 
associated fluctuations anomalously time-correlated? Can nonmarkovian processes be present when the 
system is nonextensive (i.e., when $0 \le \alpha/d <1$)? Plenty of intriguing questions that, sooner or 
later, will have to be answered, mainly through computational work (at least the first approaches). Better 
sooner than later!  

\subsection{Optimization techniques; Simulated annealing}

The so called {\it Optimization problem} consists basically in determining  the {\it global} minimum (or 
minima, if degeneracy is present) of a given cost funtion $E(x)$, where $x$ is a discrete or continuous 
$d$-dimensional variable. This problem  can become extremely complex depending on the cost function 
having a large number of {\it local} minima, and on the dimension $d$ being high. For the ubiquitous 
cases (in physics, chemistry, neural networks, engineering, finances, etc) for which analytic discussion is 
not tractable, a variety of computational algorithms have been developed.  A special place among these is 
occupied, because of its efficiency and paradigmatic value, by the {\it Simulated Annealing} (SA) 
introduced in 1983 by Kirkpatrick et al\cite{kirkpatrick}. Its denomination comes from its total analogy 
with the well known annealing technique, frequently used in Metallurgy for making a molten metal to 
reach its crystalline state (global minimum of the relevant thermodynamic energy). In SA, one or more 
artificial temperatures are introduced and gradually cooled, acting as a source of stochasticity, extremely 
convenient for eventually detrapping from local minima. Near the end of the process, the system 
hopefully is in the  attractive basin of one of the global minima, the temperature is practically zero, and 
the algorithm asymptotically becomes a steepest descent one. The challenge is to cool the temperature 
the quickest we can {\it but} still having the guarantee that no definitive trapping in any local minimum 
will occur. More precisely speaking, we search for the quickest annealing (i.e., in some sense 
approaching a {\it quenching}) which preserves the probability of ending in a global minimum being 
equal to one. SA strictly follows a BG scheme. Let us illustrate for continuous $x$. The system ``tries" to 
visit, according to a {\it visiting distribution} assumed to be Gaussian in the neighborhood of its actual 
state. The jump is always accepted if it is ``downhill", i.e., if the cost function decreases. If it is ``uphill", 
the jump {\it might} be accepted with a probability given by the Boltzmann factor corresponding to that 
cost function. It has been shown that the probability of ending on the global minimum equals unity if 
$T(t)$ decreases logarithmically with time $t$. This algorithm is sometimes referred to as {\it Classical 
Simulated Annealing} (CSA) or Boltzmann machine. We easily recognize that, if instead of decreasing, 
the temperature was maintained fixed, this procedure precisely would be the well known Metropolis et al 
one for simulating BG thermostatistical equilibrium.

This optimization machine has been generalized within the present statistics as follows\cite{tsasta}. The 
visiting ditribution is generalized to be a $q_V$-Gaussian, and the acceptance Boltzmann factor is 
generalized to be a $q_A$-generalized factor, where $q_V$ and $q_A$ respectively are the {\it visiting} 
and {\it acceptance} entropic index. The cooling schedule is generalized as follows:
\begin{equation}
T(t) =  T(1)\; \frac{\ln_q\;[1/2]}{\ln_q\;[1/(t+1)]}\;\;\;\;\;(t=1,\;2,\;3,\;...)
\end{equation}
This is the {\it  Generalized Simulated Annealing}. This machine is characterized by $(q_V, q_A)$. The 
choice $(1,1)$ corresponds to CSA, and the choice $(2,1)$ corresponds to the so called {\it Fast 
Simulated Annealing} (FSA). The CSA corresponds to a cooling given by
\begin{equation}
T(t) =  T(1)\; \frac{\ln 2}{\ln (1+t)}\;\;\;\;\;(t=1,\;2,\;3,\;...)
\end{equation}
The FSA corresponds to a faster cooling given by
\begin{equation}
T(t) =  T(1)\; \frac{1}{t}\;\;\;\;\;(t=1,\;2,\;3,\;...)
\end{equation}
The limiting case $q_V=3$ corresponds to
\begin{equation}
T(t) =  T(1)\; \frac{3}{(t+1)^2-1}\;\;\;\;\;(t=1,\;2,\;3,\;...)
\end{equation}
These particular cases illustrate the great computational advantage that can be obtained by speeding up 
the algorithm by conveniently choosing $q_A$ (see also \cite{nishimori}). In practice, a convenient 
choice for $q_A$ is slightly below 3. The choice of $q_V$ seems to be more model-dependent. Details 
can be seen in a by now vast literature\cite{penna,straub,schulte,okamoto,serra,xiang,lemes,kleber}, in 
which applications have been done and variations have been performed concerning a variety of classical 
and  quantum physical problems, the Traveling Salesman Problem, and many others. The first 
application\cite{muntsa} in quantum chemistry concerned simple molecules of the series $CH_ 3-R$ and 
some others, including the $H_20_3$ one, by using the MOPAC program package. Nowadays, Straub (in 
Boston), Okamoto (in Okazaki), and Ellis-Mundim-Bisch (respectively in Chicago, Salvador and Rio de 
Janeiro) and co-workers are currently improving and applying these techniques to complex molecules 
such as polypeptides, in particular with the aim of studying the important, though hard, {\it protein 
folding} problem.

\section{FINAL REMARKS}

Boltzmann-Gibbs statistical mechanics and standard thermodynamics do not seem to be universal. They 
have domains of applicability quite poorly known nowadays. The precise knowledge of the restrictions 
for their validity is conceptually and practically very important. A nonextensive generalization of these 
formalisms is now available\cite{tsallis,curado}; see Table I. It has been developed to cover at least some 
of these difficulties. Several important types of systems have been focused on in the present paper, which 
should substantially clarify the situation. These efforts span a wide epistemological variety, which goes 
from clear-cut theories to phenomenological ones, to quite well or less well understood fittings and 
connections. As exhibited at length here, the areas on which this formalism has been satisfactorily 
applied includes physics, astronomy, chemistry, mathematics, biology, economics, linguistics, cognitive 
psychology, etc. However, in spite of all this sensible progress, some inter-related crucial points are still 
to be understood and established on a neat and transparent basis. These include (i) the zeroth principle of 
thermodynamics and its connections with the thermodynamic limit, properties which would in principle 
exhibit the mathematical connection of the {\it weak} violation of the BG statistics (i.e., introduction of 
$\tilde{N}$ within the $q=1$ formalism) with the {\it strong} one ($q \ne 1$); (ii) the functional 
dependence of $q$ on $(\alpha,d)$ for long-range interacting Hamiltonians ($0 \le \alpha/d \le 1$) and its 
connection with anomalous diffusion; (iii) the physical interpretation of the $\{p_i\}$ distribution and of 
the escort one $\{P_i\}$, as well as clear-cut prescriptions for using one or the other when fitting 
experimental data (in the meanwhile, it appears that $\{P_i\}$ is the one to be used for equilibrium 
distributions of kinetic, potential or total energies, whereas $\{p_i\}$ is the one to  be used for real-space 
diffusive nonequlibrium phenomena, either Levy or Student's\cite{student} distributions according to 
whether correlations between jumps are suspected to be absent or present); (iv) what are the generic 
physical conditions for using the microcanonical, canonical and grand-canonical ensembles, under what 
exact and fully specified conditions they are expected to be thermodynamically equivalent, and the 
possible relevance for the so called thermogravitational catastrophe; (v) the clear connection with 
microscopic dynamic properties such as (partial) lack of ergodicity and mixing, the generalization (to 
weak chaos) of the Pesin inequality, the complete domain of validity of scaling relations connecting $q$ 
to multifractality, and the possible relevance for SOC, spin-glasses and similar phenomena; (vi) the clear 
physical connection with quantum groups and, in general, deformations of relevant Lie algebras, and 
through these, the possible relevance for quantum gravity and the deep (possibly discrete, multifractal-
like) structure of space-time.

On speculative grounds, one might think of two conjectures, to be clarified (i.e., rigorously formulated), 
confirmed or refuted. The first of these conjectures can somehow (on intuitive grounds) be formulated as 
follows. {\it Strongly} ergodic/mixing phenomena are ubiquitous in Nature; essentially, they are driven 
by microscopic interactions which are short-ranged in space-time (short-range forces, short-range 
memory, nonfractal boundary conditions); their basic geometry tends to be continuous, Euclidean-like; 
their thermodynamics is extensive; their central laws (energy distribution at equilibrium, time-relaxation 
towards equilibrium) generically are exponentials;  and their thermostatistical foundation is  Boltzmann-
Gibbs statistics (i.e., $q=1$). But {\it weakly} ergodic/mixing phenomena also are ubiquitous in Nature 
(e.g., biological, socio-economical, human cognitive phenomena, etc); essentially, they are driven by 
microscopic interactions which are long-ranged in space time (long-range forces, nonmarkovian memory, 
fractal boundary conditions); their basic geometry tends to be discrete, multifractal-like; their 
thermodynamics is nonextensive; their central laws (energy distribution at equilibrium, time-relaxation 
towards equilibrium) generically are power-laws; and the thermostatistical foundation of (at least some 
of) them (hopefully !) is the $q \ne 1$ statistics. The allowance for nonextensivity, in general, and for a 
nonextensive entropy, in particular, appears to be the ``price" to be payed  in order that Boltzmann's 
``mechanical" (i.e., {\it one} system evolving along time) manner of thinking about macroscopic systems 
coincides, at the level of the concrete mathematical results to be compared with the experimental data, 
with Gibbs' ``ensemble" (i.e., {\it many} systems at a fixed time) manner of thinking. This coincidence of 
results is, since one century, well known and understood for standard sytems. Our aim here is to extend it 
to a large variety of anomalous systems.

The second of these conjectures is, at the present moment, so hard to rationalize that I dare to mention it 
here {\it only} because, after having been exposed to so many mathematical and physical arguments (that 
have been included in the present review), the reader might accept to honor me with his (her) indulgence, 
and have a look at the following few, intuitive lines. I believe that a deep analogy (maybe a kind of 
isomorphism, through the use of mathematical structures like the co-homology groups) exists between 
crystallographic structures such as crystal - quasicrystal - fluid, and nonlinear dynamics such as 
integrable - (weak) chaotic - (strong) chaotic. In some sense, they appear as space and time versions of 
the same mathematical structures. The first case concerns crystals (i.e., $d$-dimensional Bravais lattices) 
and integrable dynamics (i.e., motion on simple orbits), and its essential invariance is the {\it discrete 
translational} one. The third case concerns strongly disordered systems like fluids (liquid, gases) and 
strongly chaotic dynamics, and its essential invariance is the {\it continuous translational} one. Finally, 
the second, and intermediate, case is by far the most subtle one (and probably this is why it is the one that 
humanity took the longest time to discover), and concerns quasicrystals (e.g., Penrose tilings, amorphous 
substances like glasses, spin-glasses, and other structures known to have (multi)fractal scalings; probably 
most of the so called complex spatial
phenomena belong to this group) and weakly ergodic dynamics (e.g., edge of chaos, strange attractors, 
self-organized criticality, probably most of the so called complex time phenomena); its essential 
invariance is the {\it dilatation} one. In the first case we have the (space or time) {\it highest 
predictability}, and statistical methods are out of place. In the third case we have the (space or time) {\it 
lowest predictability}, and statistical methods exhibit their full power. Finally, in the second case, we 
have an {\it intermediate predictability}, and the statistical methods have to be ``intrinsically nonlinear" 
in some sense, in order to be applicable and useful. There will be no surprise for the reader if, at this 
point,  I admit that I believe that the statistical mechanics to be associated with the third present case of 
course is the BG one, whereas it might be the $q \ne 1$ statistical mechanics the one to be associated 
with the present second case ! (see also \cite{alemanyraul}, where some preliminary, but nevertheless 
concrete, calculations exhibit this kind of  connections). Let us now remind that Wiles' 1995 celebrated 
proof\cite{wiles} of Fermat's last theorem was deeply related to quasicrystals since it was based on the 
proof of the Taniyama-Shimura conjecture about modular elliptic curves and used certain Hecke 
algebras. Consequently, I hardly dare to explicitly state a simple and unavoidable corollary, namely that, 
if my present second conjecture turns out to be, in some nontrivial and precise sense, correct, then the $q 
\ne 1$ statistical mechanics must be related to Fermat's last theorem !

Through the complete analysis, in more detailed terms, of the various aspects tackled in the present 
review, we could learn a lot and, very especially, {\it (precisely) when} the celebrated Boltzmann factor 
is the correct theoretical description of natural systems at thermal equilibrium ! This famous and so 
useful factor would then become, not a ``dogma", as referred to by Takens\cite{takens}, but a theorem !

We acknowledge extremely fruitful discussions, along the years, with E.M.F. Curado, A. Plastino, A.R. 
Plastino, R. Maynard, T.A. Kaplan,  S.D. Mahanti, P.M. Duxbury, A. Overhauser, A.C.N. de Magalhaes, 
R.S. Mendes, A.K. Rajagopal, P. Quarati, L. Borland, C. Anteneodo, D.A. Stariolo, F.A. Tamarit, S.A. 
Cannas, S. Abe, B.M. Boghosian, M.L. Lyra, T.J.J. Penna, H.J. Herrmann, F.C. Alcaraz, A. Coniglio, I. 
Procaccia, J.-P. Eckmann, E.P. Borges, D. Prato, A. Craievich and so many others that it would be an 
almost impossible task to name them all. Also, I am grateful to N. Caticha for having drawn my attention 
onto the simulated annealing technique and the plausibility of using nonextensive statistics in order to 
improve it. Finally, I am very grateful to C. Anteneodo, H.N. Nazareno, P.E. de Brito, A. Upadhyaya, 
J.P. Rieu, J.A. Glazier, Y. Sawada,  F.M. Ramos, R.R. Rosa, C. Rodrigues Neto, I. Bediaga, E.M.F. 
Curado, J. Miranda and G. Guerberoff for making their results available to me prior to publication. This 
work was partially supported by CNPq and PRONEX/FINEP (Brazilian Agencies).

{\bf FIGURE CAPTIONS}

{\bf Figure 1} - $W=2$ illustration of the {\it escort} probabilities: $P^{(q)}=\frac{p^q}{p^q+(1-p)^q}$.

{\bf Figure 2} - Generalization (Eq. (42)) of the Boltzmann factor (recovered for $q=1$) as function of 
the energy $E$ at a given renormalized temperature $T^{\prime}$,  assuming a constant density of 
states. From top to bottom at low energies: $q=0,\;1/4,\;1/2,\;2/3,\;1,\;3,\;\infty$ (the vertical line at 
$E/T^{\prime}=1$ belongs to the limiting $q=0$ distribution; the $q \rightarrow \infty$ distribution 
collapses on the ordinate). All $q>1$ curves have a $(T^{\prime}/E)^{q/(q-1)}$ tail; all $q<1$ curves 
have a cut-off at $ E/T^{\prime}=1/(1-q)$.       

{\bf Figure 3} - Log-log plot of some cases like those of Fig. 2 ($T^{\prime}=1,\;5$ for each value of 
$q$).     

{\bf Figure 4} - Central conjecture of the present work, assuming a Hamiltonian system  which includes 
two-body (attractive) interactions which, at long distances, decay as $r^{-\alpha}$. The crossover at 
$t=\tau$ is expected to be slower than indicated in the figure (for space reasons).   

{\bf Figure 5} - The one-jump distributions $p_q(x)$ for typical values of $q$. The $q \rightarrow -
\infty$ distribution is the uniform one in the interval $[-1,1]$; $q=1$ and $q=2$ respectively correspond 
to Gaussian and Lorentzian distributions; the $q \rightarrow 3$ is completely flat. For $q<1$ there is a 
cut-off at $|x|/\sigma = [(3-q)/(1-q)]^{1/2}$.       

{\bf Figure 6} - The $q$-dependence of the dimensionless diffusion coefficient $\Delta_q$ ({\it width} of 
the properly scaled distribution $p_q(x,N)$ in the limit $N \rightarrow \infty$). In the limits $q 
\rightarrow 5/3 - 0$ and $q \rightarrow 5/3+0$ we respectively have $\Delta_q \sim  [4/9]/[(5/3)-q]$ and 
$\Delta_q \sim [4/(9\pi^{1/2}]/[q-(5/3)]$; also, $\lim_{q \rightarrow 3} \Delta_q=2/\pi^{1/2}$.  

{\bf Figure 7} - The index $q$ to be associated with a person whose critical value corresponding to Eq. 
(69) is $V_c$. People with $q<1$ ($q>1$) tend to avoid (seek) risks for that particular game. The case 
$q=1$ corresponds to an ideally rational agent. 

{\bf Figure 8} - Distribution of the transverse momenta $p_T$ obtained in electron-positron frontal 
collisions of energy $W$ varying from $14$ to $161\;Gev$. The dotted line corresponds to $q=1$ (i.e., a 
Hagedorn type of fitting as given by Eq. (87)) for all values of $W$. The solid lines correspond to $q \ne 
1$ fittings. 

{\bf Figure 9} - The values of $q$ and $T_0$ used in the fittings of Fig. 8. When $W$ approaches zero, 
$q$ approaches unity, i.e., Hagedorn's theory; $T_0$ is essentially insensitive to $W$, as physically 
desirable.

{\bf Figure 10} - Time evolution of $\xi \equiv N(t)/N(0)$ associated with $MbCO$ in glycerol-water. 
Dots: experimental data. Dashed lines: fittings with Frauenfelder's empiric law (Eq. (88) or Eq. (90)). 
Solid lines: fittings with the solutions of Eq. (91) (see Fig. 11).

{\bf Figure 11} - Temperature dependences of the parameters used to fit the experimental data of Fig. 10.

{\bf Figure 12} - Distribution of the ``vertical" velocities during diffusion of endodermal Hydra cells in 
an ectodermal aggregate. The abcissa units are $10^{-6}\;m/hour$. The fitting was obtained using 
$q=1.53$ (see the text).

{\bf Figure 13} - Distribution of ISI and PRD papers having received $x$ citations. $(a)$ and $(b)$ 
exhibit the fittings in \cite{redner}; $( c)$ and $(d)$ exhibit our present fittings (see the text). 

{\bf Figure 14} - Distributions of price changes for US dollar-German mark exchange rates and fittings 
using assymetric $q$-distributions (see the text).

{\bf Figure 15} - Distributions of velocity differences and fitting using assymetric $q$-distributions (see 
the text).

\newpage

\begin{table}

\caption{Some useful formulae written, through Eqs. (9) and (10), in a Boltzmann-Gibbs-like form}
\label{sap}
\begin{center}
\begin{tabular}{|l|l|} \hline 
Equiprobability entropy    & $S_q = k \ln_q\;W$ \\ \hline
Generic entropy  & $S_q= -k \langle \ln_q \rho \rangle_q$ \\ \hline 
Canonical equilibrium distribution  & $\rho_q=\frac{e_q^{-\beta ({\cal{H}}-U_q) / Tr \rho_{q}^q}}{ 
Tr\; e_q^{-\beta ({\cal{H}}-U_q) / Tr \rho_{q}^q }}   =\frac{
e_q^{-\beta^{\prime}\cal{H}}}{Tr e_q^{-\beta^{\prime}\cal{H}}}\; (\beta^{\prime} \equiv 
\frac{\beta}{Tr \rho_q^q+(1-q) \beta U_q}$)
\\  \hline
Partition functions  &  $\bar{Z}_q =Tr\;e_q^{-\beta ({\cal{H}}-U_q) / Tr \rho_{q}^q}\;\;(\ln_q Z_q= 
\ln_q \bar{Z}_q - \beta U_q) $ \\   \hline
Internal energy  &  $U_q=-\frac{\partial }{\partial \beta}   \ln_q Z_q$\\  \hline
Free energy  &   $F_q=U_q-TS_q=-\frac{1}{\beta}\ln_q Z_q$\\  \hline
Anomalous diffusion probability distribution   &  $p_q(x) = \frac{e_q^{- \beta x^2}}{\int dy\; e_q^{- 
\beta y^2}}$ \\   \hline
Sensitivity to the initial conditions ($d=1$)  &  $ \lim_{\Delta x(0) \rightarrow 0} \frac{\Delta 
x(t)}{\Delta x(0)} = e_q^{\lambda_q\;t}      $ \\   \hline
Likelihood function   &  $  W_q(\{p_i\}) \propto  e_q^{S_q(\{p_i\})}          $ \\  \hline
Power-law interactions ($\propto R^{-\alpha}$)  &$ \frac{U(N,T)}{N\tilde{N}} \sim 
u(\frac{T}{\tilde{N}})\;\;(\tilde{N} \equiv N^*[1+ (\alpha/d)];\;N^* \equiv 
\ln_{[\alpha/d]}N)\;\;\;\;\;\;\;\;\;$ \\  \hline
Simulated annealing (cooling rythm)  &  $\frac{T(t)}{T(1)}= \frac{\ln_q\;[1/2]}{\ln_q\;[1/(t+1)]}$ \\ 
\hline
\end{tabular}
\end{center}
\end{table}

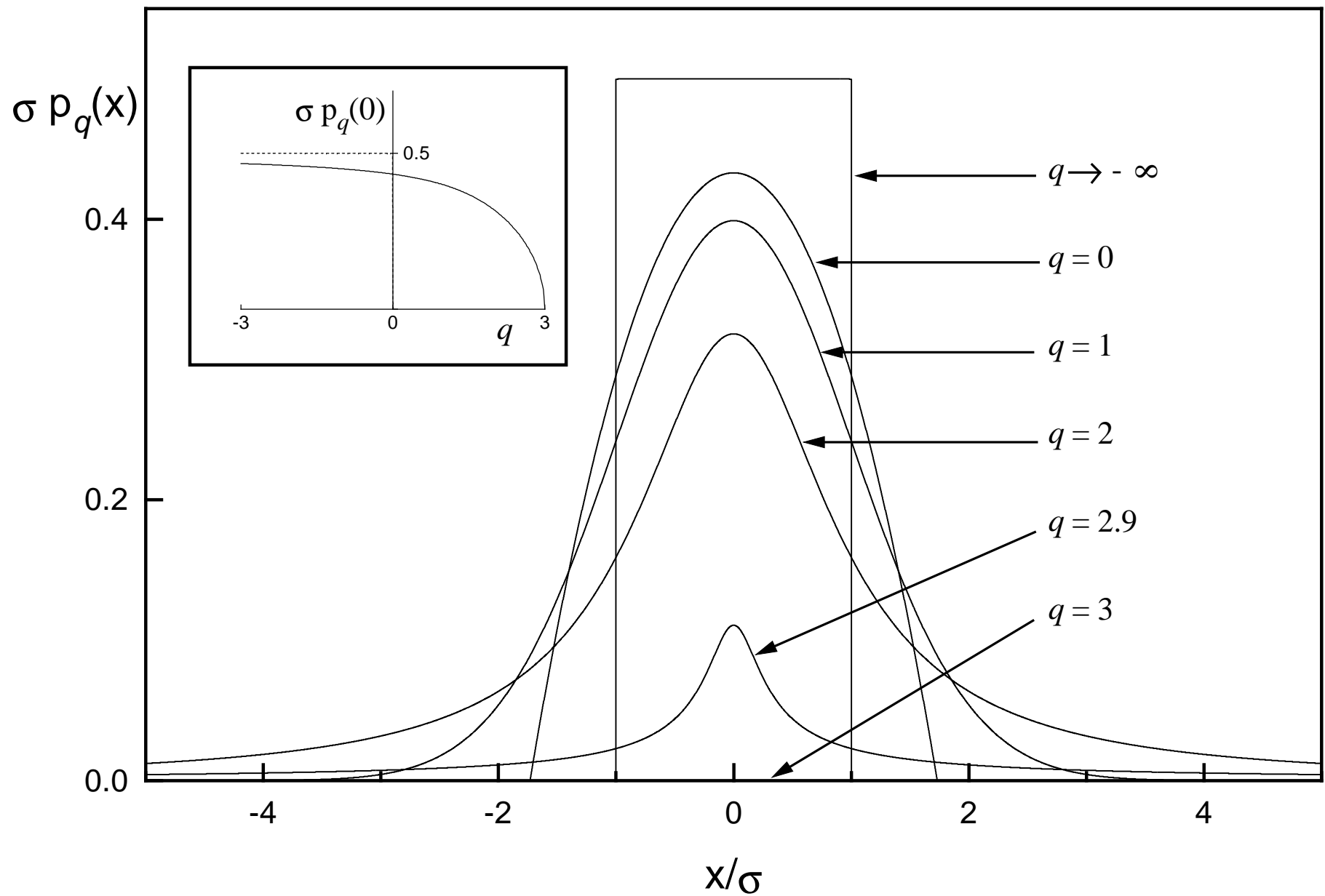
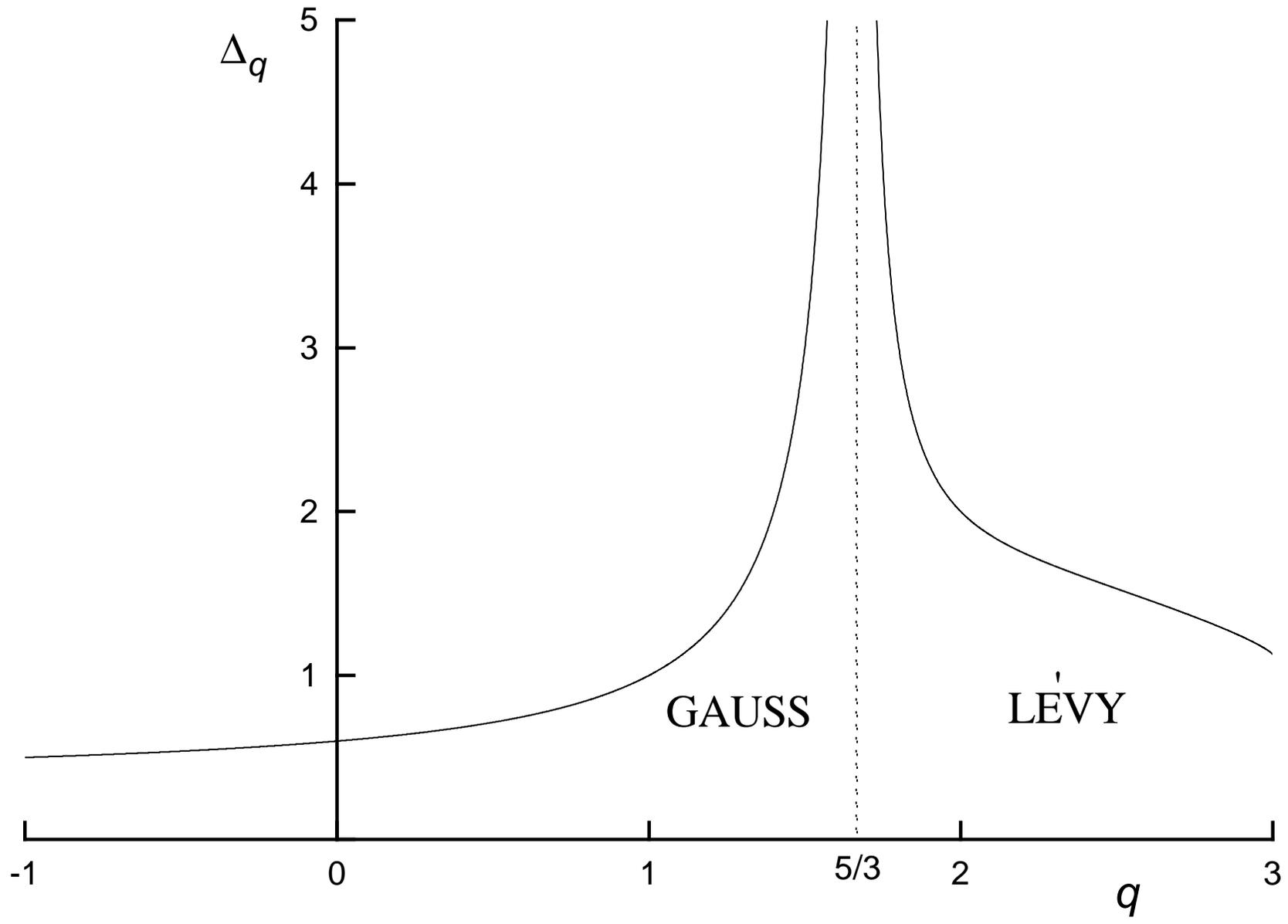
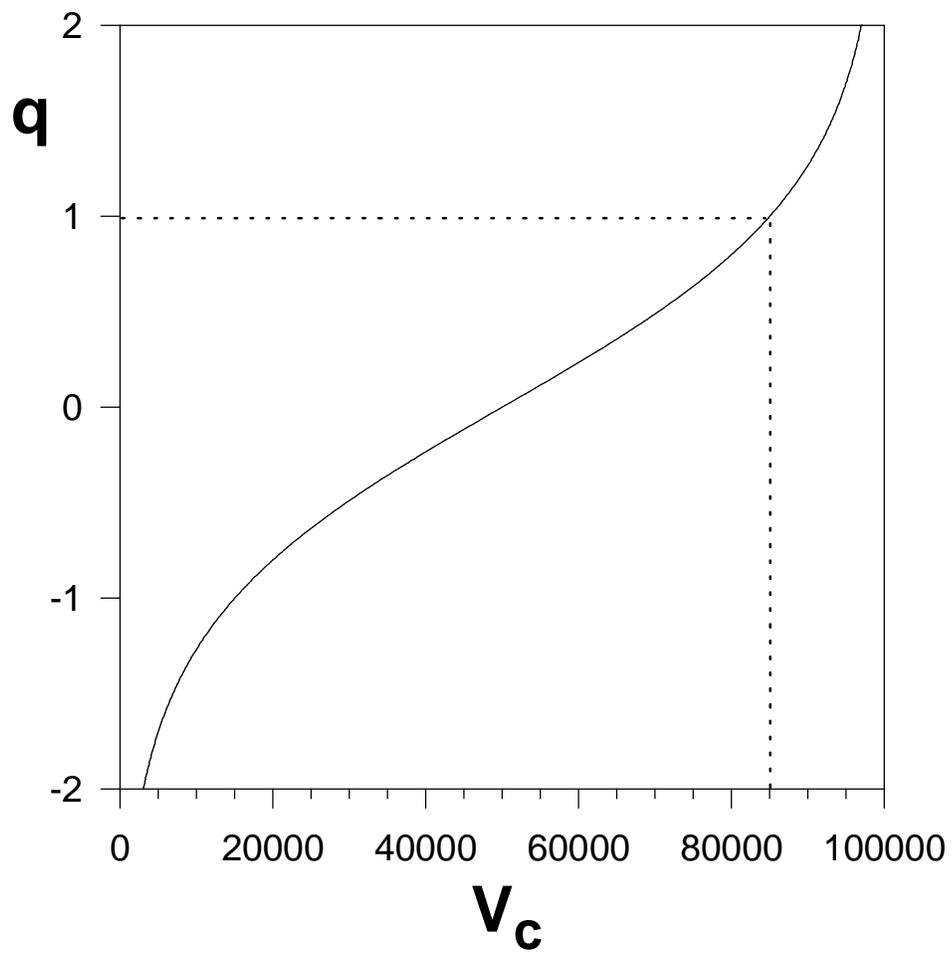
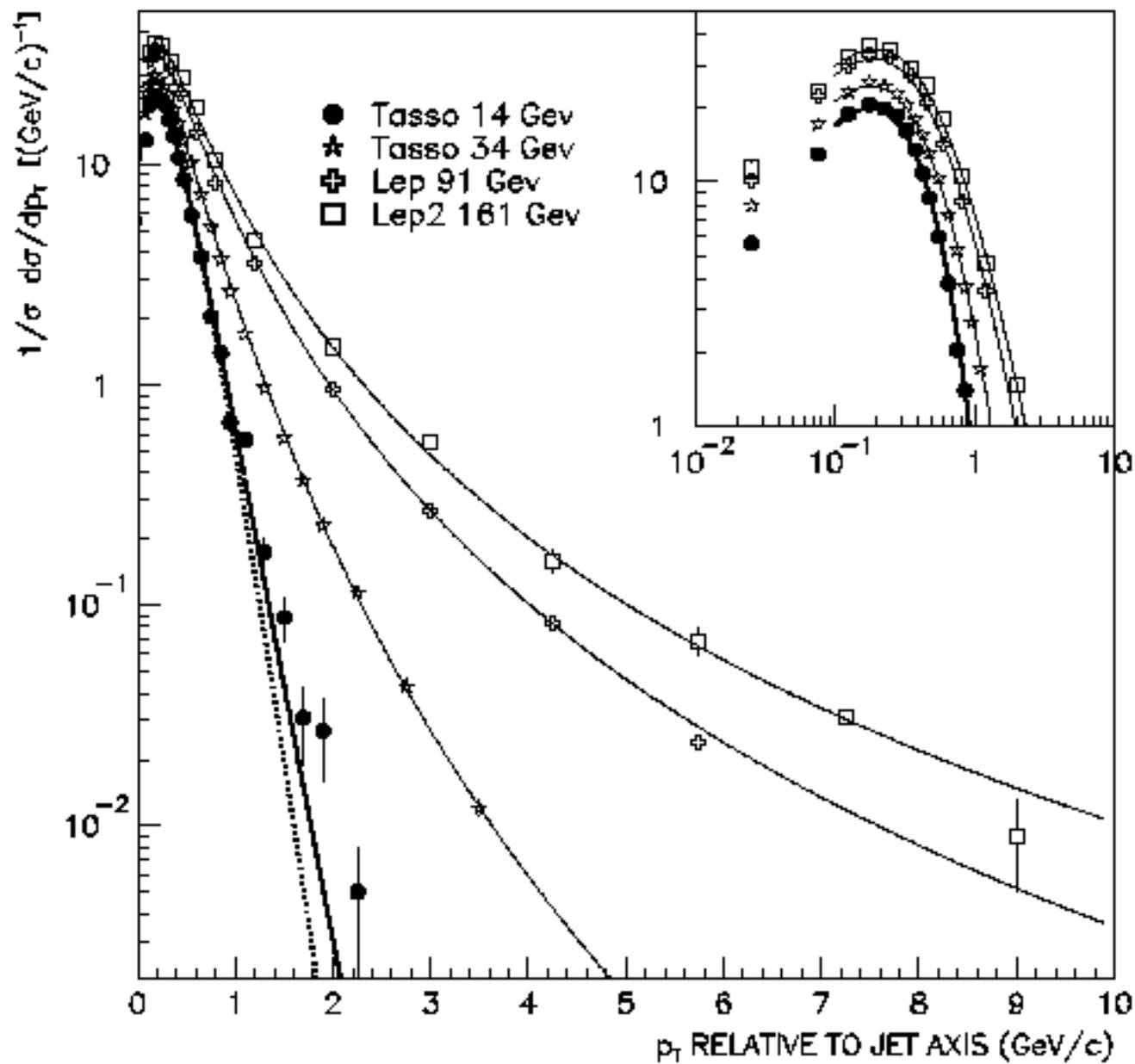
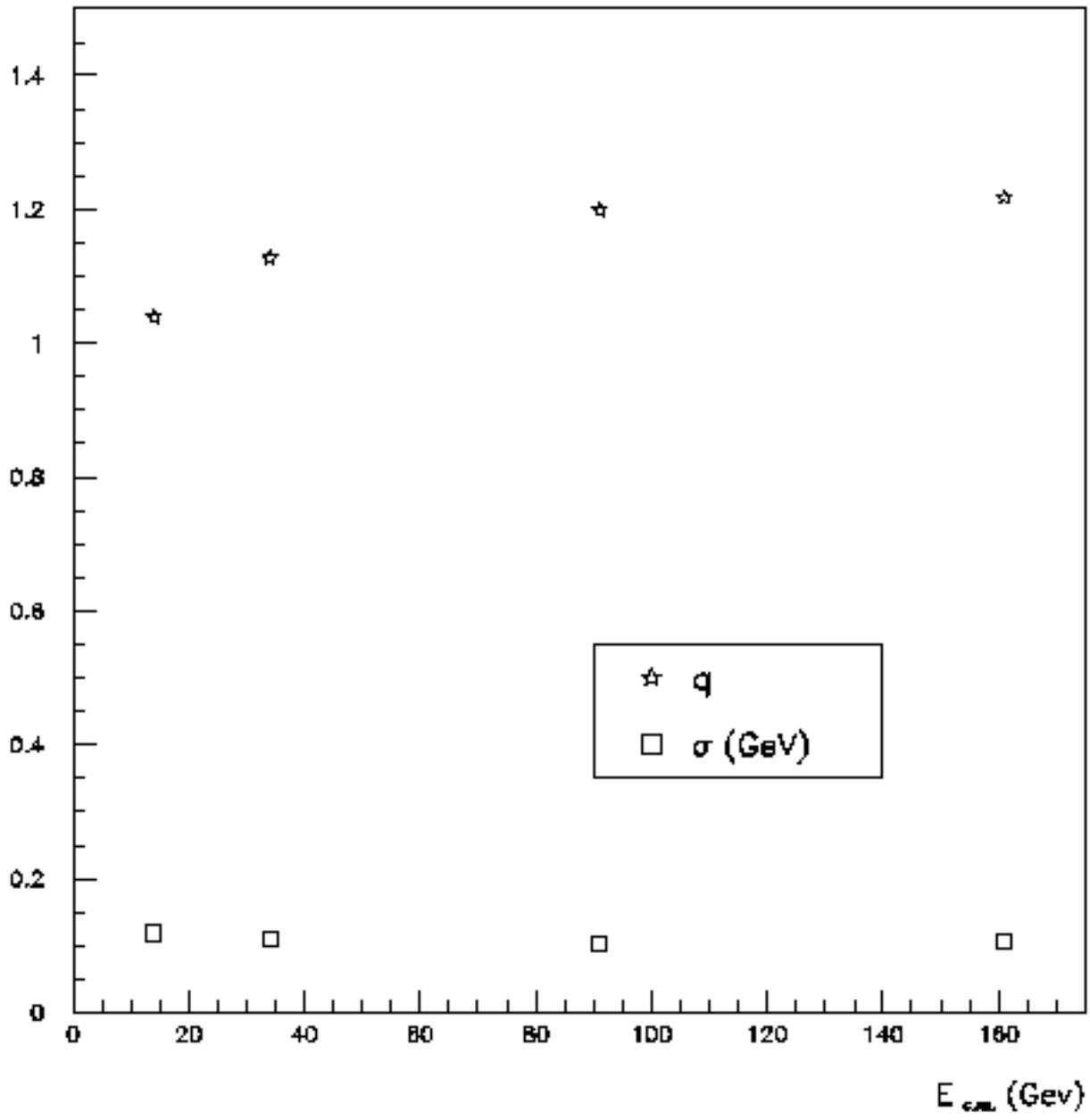
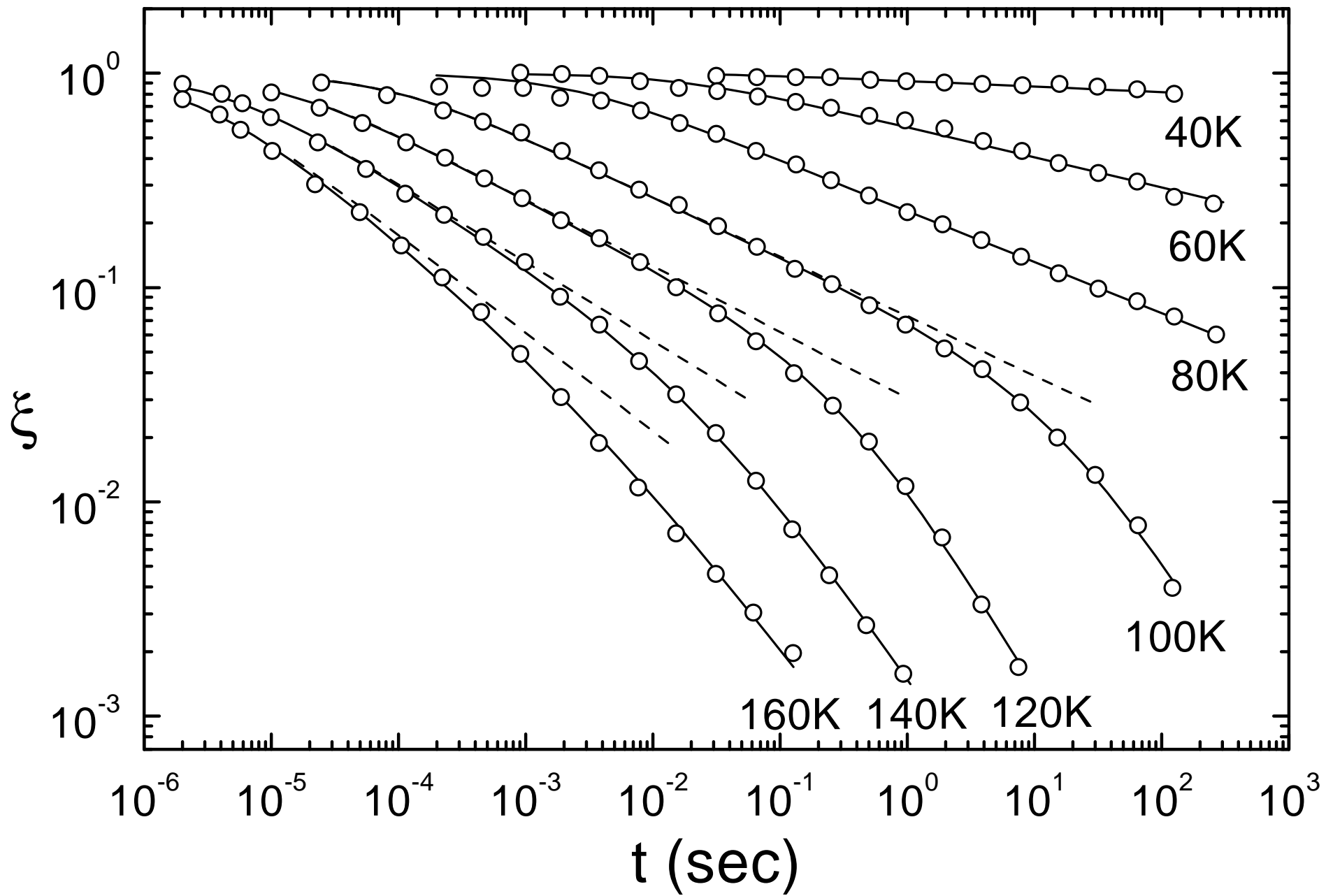
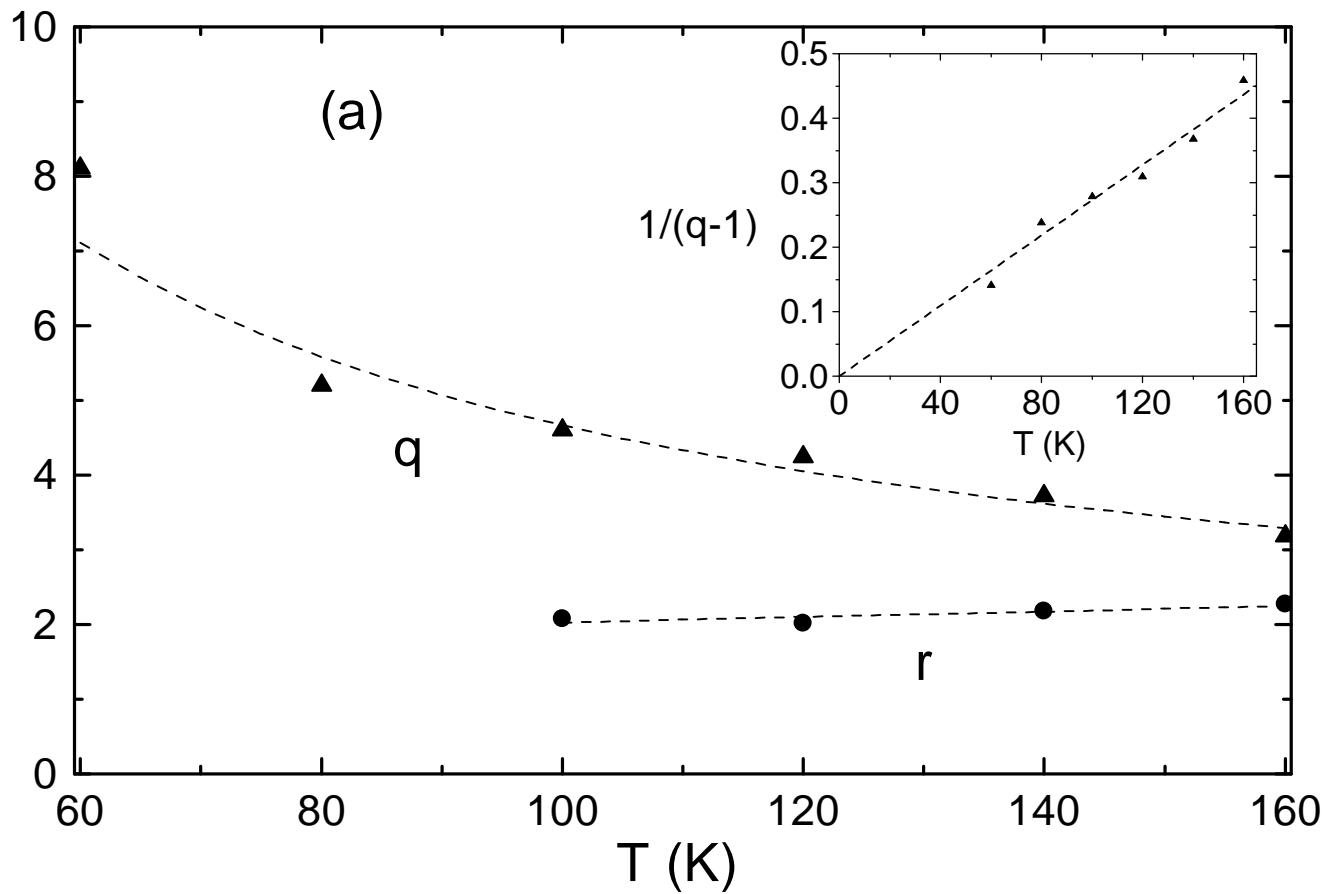
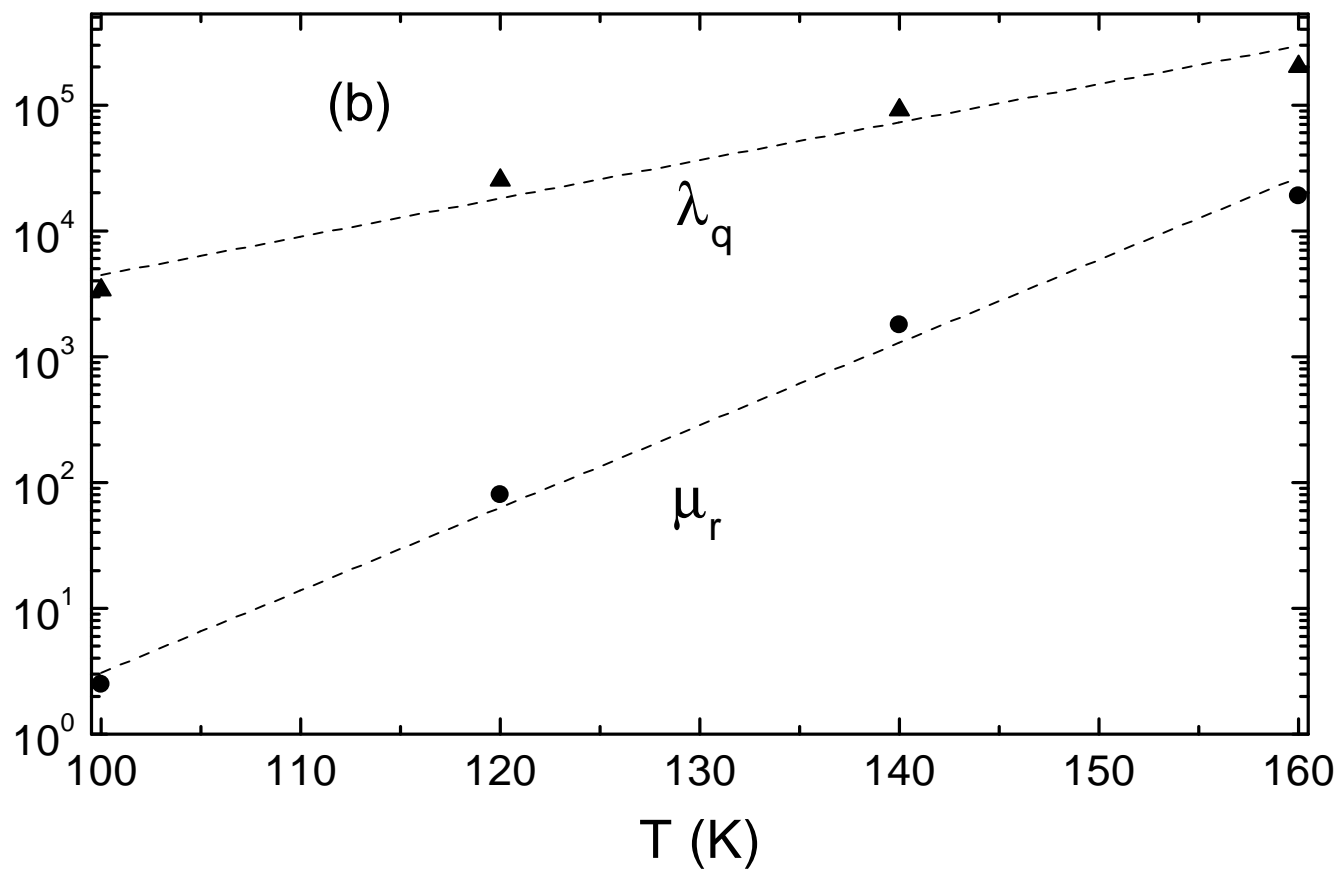
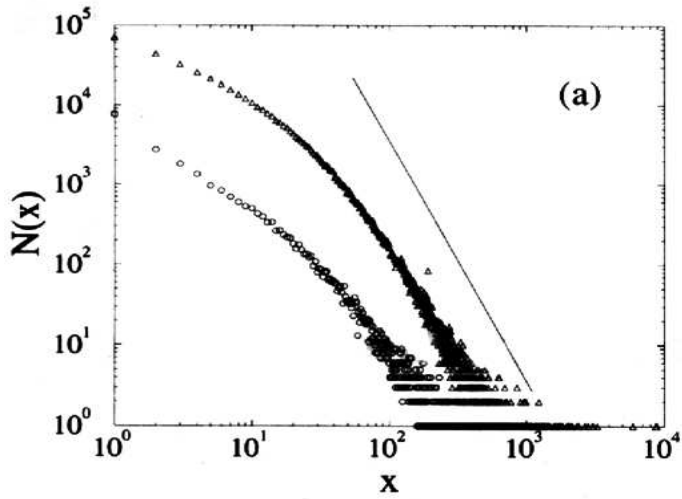
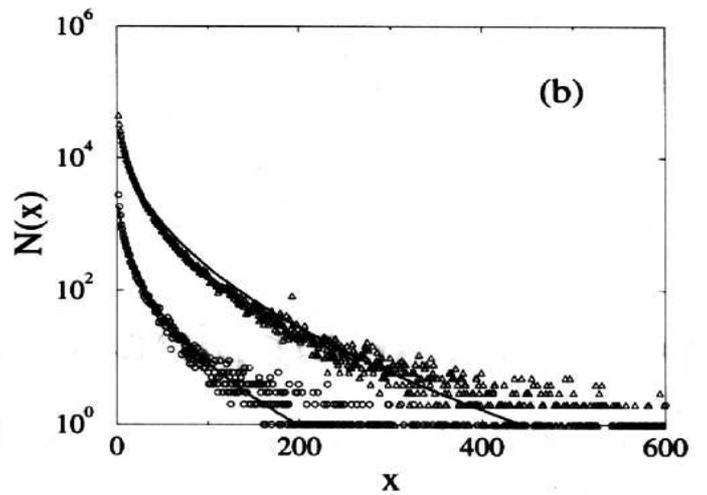
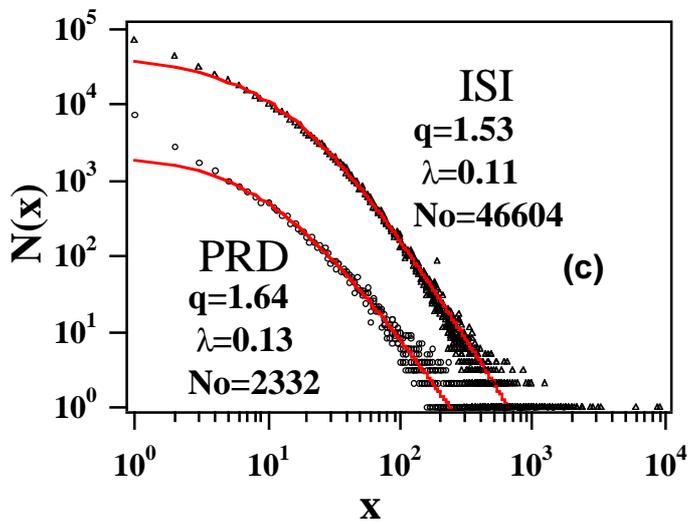
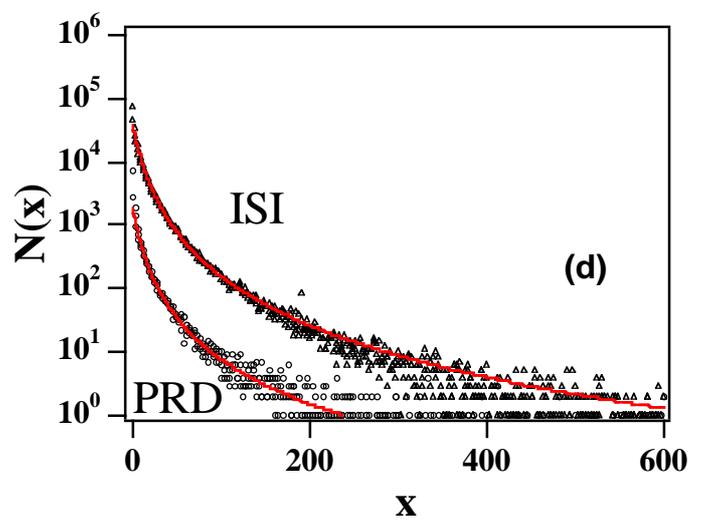
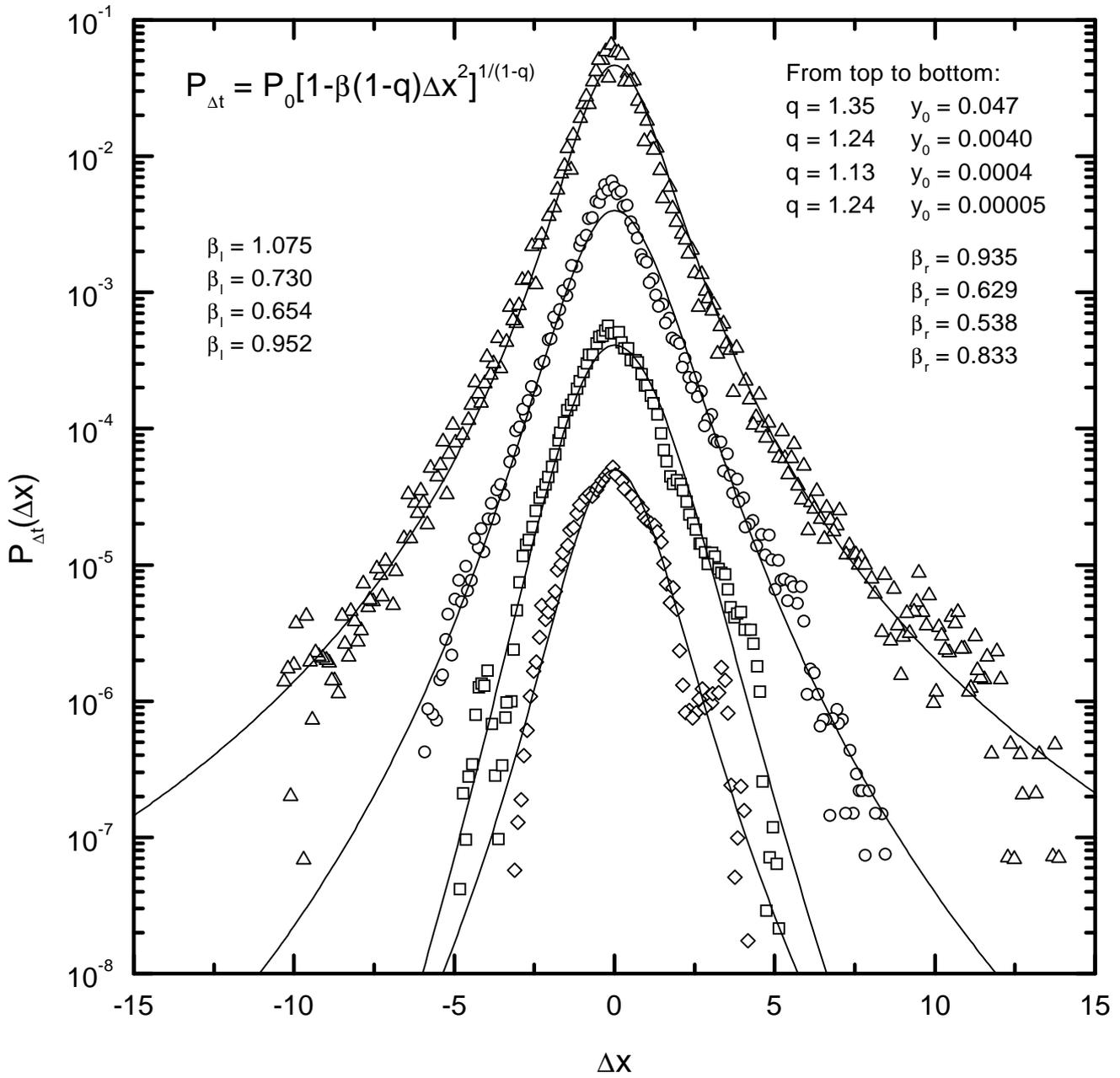
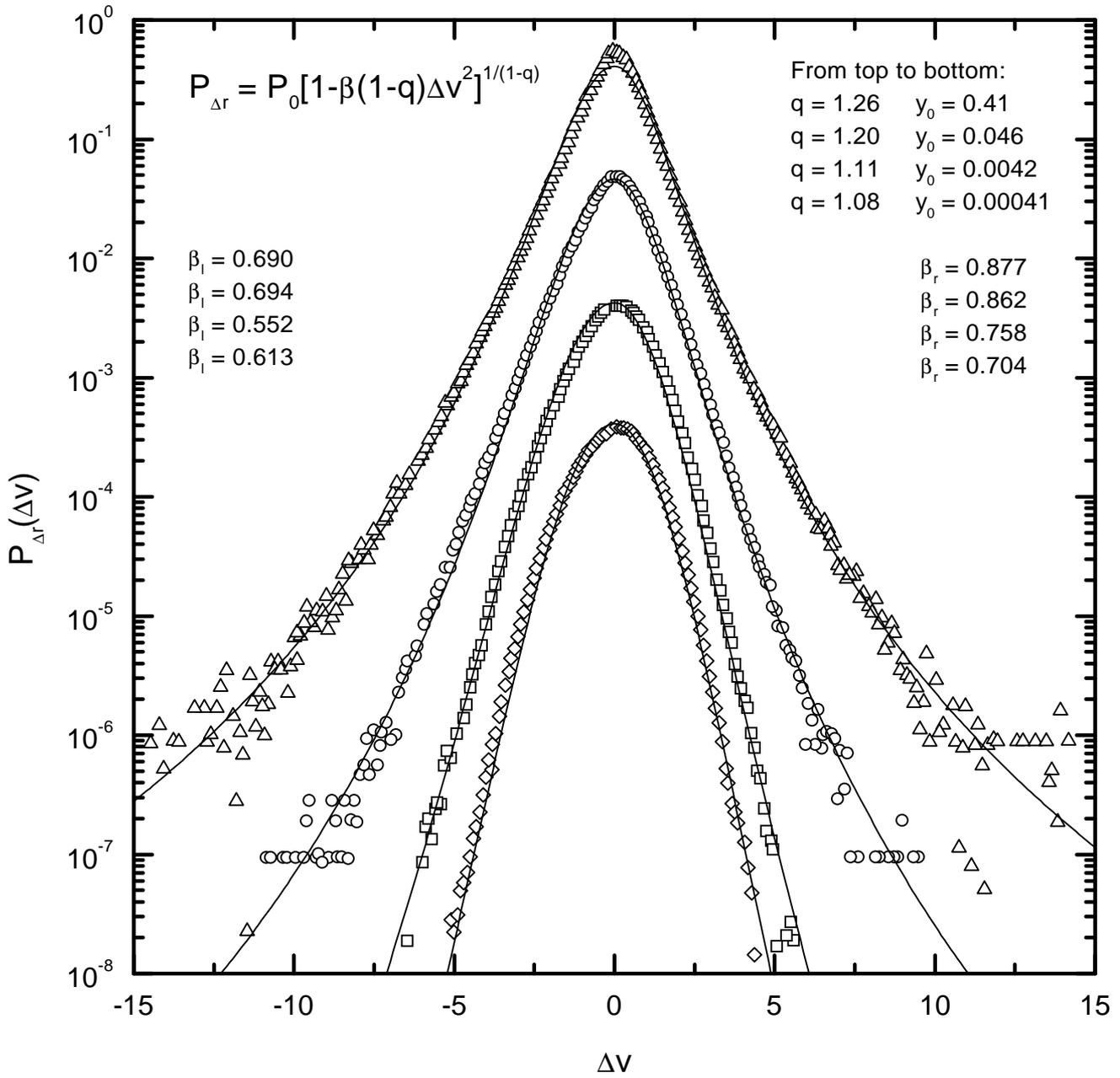
\end{document}